\documentclass[useAMS,usenatbib]{mn2e}
\usepackage[a4paper,top=20mm,bottom=20mm, left=15mm, right=15mm]{geometry}
\usepackage{amssymb,amsmath}
\usepackage{graphicx}
\usepackage{float}
\usepackage{afterpage}
\usepackage[toc,page]{appendix}
\usepackage{color}
\usepackage{multirow}
\usepackage{url}
\usepackage{xcolor}
\usepackage{xspace}
\usepackage{dblfloatfix}
\usepackage{comment}
\usepackage{booktabs}
\usepackage{chngpage}

\usepackage{aas_macros}

\usepackage{lineno}

\def\reff@jnl#1{{\rm#1\/}}
\def\aj{\reff@jnl{AJ}}         
\def\araa{\reff@jnl{ARA\&A}}      
\def\apj{\reff@jnl{ApJ}}        
\def\apjl{\reff@jnl{ApJ}}        
\def\apjs{\reff@jnl{ApJS}}       
\def\aap{\reff@jnl{A\&A}}        
\def\aapr{\reff@jnl{A\&A~Rev.}}     
\def\aaps{\reff@jnl{A\&AS}}       
\def\mnras{\reff@jnl{MNRAS}}      
\def\physrep{\reff@jnl{Physics Reports}}
\def\prd{\reff@jnl{Phys.Rev.D}}     
\def\prl{\reff@jnl{Phys.Rev.Lett}}   
\def\pasp{\reff@jnl{PASP}}       
\def\pasj{\reff@jnl{PASJ}}       
\def\nat{\reff@jnl{Nature}}       
\def\jcap{\reff@jnl{JCAP}}   
\def\memsai{\reff@jnl{MemSAI}} 
\def\na{\reff@jnl{New Astronomy}}       
\def\procspie{\reff@jnl{SPIE}}       
\def\pasa{\reff@jnl{PASA}}       

\def\Sref#1{Sec.~\ref{#1}\xspace}
\def\Fref#1{Fig.~\ref{#1}\xspace}
\def\Tref#1{Table~\ref{#1}\xspace}

\title[{Spectroscopic incompleteness}]{The impact of spectroscopic incompleteness in direct calibration of redshift distributions for weak lensing surveys} 
\author[Hartley, Chang and the DES Collaboration]{
\parbox{\textwidth}{
\Large
W.~G.~Hartley,$^{1,2,3, *}$
C.~Chang,$^{4,5}$ 
S.~Samani,$^{1}$ 
A.~Carnero~Rosell,$^{6}$ 
T.~M.~Davis,$^{7}$
B.~Hoyle,$^{8,9}$
D.~Gruen,$^{10,11,12}$
J.~Asorey,$^{6}$
J.~Gschwend,$^{13,14}$
C.~Lidman,$^{15}$
K.~Kuehn,$^{16,17}$
A.~King,$^{7}$
M.~M.~Rau,$^{18}$
R.~H.~Wechsler,$^{10,11,12}$ 
J.~DeRose,$^{19,20}$
S.~R.~Hinton,$^{7}$
L.~Whiteway,$^{1}$
T.~M.~C.~Abbott,$^{21}$
M.~Aguena,$^{13,22}$
S.~Allam,$^{23}$
J.~Annis,$^{23}$
S.~Avila,$^{24}$
G.~M.~Bernstein,$^{25}$
E.~Bertin,$^{26,27}$
S.~L.~Bridle,$^{28}$
D.~Brooks,$^{1}$
D.~L.~Burke,$^{11,12}$
M.~Carrasco~Kind,$^{29,30}$
J.~Carretero,$^{31}$
F.~J.~Castander,$^{32,33}$
R.~Cawthon,$^{34}$
M.~Costanzi,$^{35,36}$
L.~N.~da Costa,$^{13,14}$
S.~Desai,$^{37}$
H.~T.~Diehl,$^{23}$
J.~P.~Dietrich,$^{38}$
B.~Flaugher,$^{23}$
P.~Fosalba,$^{32,33}$
J.~Frieman,$^{23,5}$
J.~Garc\'ia-Bellido,$^{24}$
E.~Gaztanaga,$^{32,33}$
D.~W.~Gerdes,$^{39,40}$
R.~A.~Gruendl,$^{29,30}$
G.~Gutierrez,$^{23}$
D.~L.~Hollowood,$^{20}$
K.~Honscheid,$^{41,42}$
D.~J.~James,$^{43}$
S.~Kent,$^{23,5}$
E.~Krause,$^{44}$
N.~Kuropatkin,$^{23}$
O.~Lahav,$^{1}$
M.~Lima,$^{22,13}$
M.~A.~G.~Maia,$^{13,14}$
J.~L.~Marshall,$^{45}$
P.~Melchior,$^{46}$
F.~Menanteau,$^{29,30}$
R.~Miquel,$^{16,47}$
R.~L.~C.~Ogando,$^{13,14}$
A.~Palmese,$^{23,5}$
F.~Paz-Chinch\'{o}n,$^{29,30}$
A.~A.~Plazas,$^{46}$
A.~Roodman,$^{11,12}$
E.~S.~Rykoff,$^{11,12}$
E.~Sanchez,$^{6}$
V.~Scarpine,$^{23}$
M.~Schubnell,$^{40}$
S.~Serrano,$^{32,33}$
I.~Sevilla-Noarbe,$^{6}$
M.~Smith,$^{48}$
M.~Soares-Santos,$^{49}$
E.~Suchyta,$^{50}$
G.~Tarle,$^{31}$
M.~A.~Troxel,$^{51}$
D.~L.~Tucker,$^{23}$
T.~N.~Varga,$^{8,9}$
J.~Weller,$^{8,9}$
and R.D.~Wilkinson$^{52}$
\begin{center} (DES Collaboration) \end{center}
}
\vspace{0.4cm}
\\
\parbox{\textwidth}{
(affiliations are listed at the end of the paper)\\
$^{*}$Corresponding author: william.hartley@unige.ch\\
}
}

\usepackage{eso-pic}

\AddToShipoutPictureBG*{%
  \AtPageUpperLeft{%
    \hspace{1.8\linewidth}%
    \raisebox{-3.5\baselineskip}{%
      \makebox[0pt][l]{\textnormal{DES-2019-0498}}
}}}%

\AddToShipoutPictureBG*{%
  \AtPageUpperLeft{%
    \hspace{1.8\linewidth}%
    \raisebox{-4.5\baselineskip}{%
      \makebox[0pt][l]{\textnormal{FERMILAB-PUB-20-106-AE}}
}}}%

\begin{document}
\date{\today}

\label{firstpage}
\maketitle

\begin{abstract}
Obtaining accurate distributions of galaxy redshifts is a critical aspect of weak lensing cosmology experiments. One of the methods used to estimate and validate redshift distributions is to apply weights to a spectroscopic sample so that their weighted photometry distribution matches the target sample. In this work we estimate the \textit{selection bias} in redshift that is introduced in this procedure. We do so by simulating the process of assembling a spectroscopic sample (including observer-assigned confidence flags) and highlight the impacts of spectroscopic target selection and redshift failures. We use the first year (Y1) weak lensing analysis in DES as an example data set but the implications generalise to all similar weak lensing surveys. We find that using colour cuts that are not available to the weak lensing galaxies can introduce biases of up to $\Delta z\sim0.04$ in the weighted mean redshift of different redshift intervals ($\Delta z\sim0.015$ in the case most relevant to DES). To assess the impact of incompleteness in spectroscopic samples, we select only objects with high observer-defined confidence flags and compare the weighted mean redshift with the true mean. We find that the mean redshift of the DES Y1 weak lensing sample is typically biased at the $\Delta z=0.005-0.05$ level after the weighting is applied. The bias we uncover can have either sign, depending on the samples and redshift interval considered. For the highest redshift bin, the bias is larger than the uncertainties in the other DES Y1 redshift calibration methods, justifying the decision of not using this method for the redshift estimations. We discuss several methods to mitigate this bias.
\end{abstract}


\section{Introduction}
\label{sec:intro}

Over the last decade, a number of new deep imaging surveys have been developed 
in order to take advantage of the cosmological information contained within the distortion of galaxy shapes by 
weak gravitational lensing. One of the quantities required to be known in order to unlock this information is the 
distribution in redshift of the galaxies whose light is being distorted. The first of the so-called stage III programmes \citep{Albrecht2006} designed to 
measure weak lensing have now been completed \citep{Heymans2013,Joudaki2017}. Current state-of-the-art 
surveys, such as the Kilo-Degree Survey \citep[KiDS,][]{deJong2015}, the Hyper SuprimeCam Survey 
\citep[HSC,][]{Aihara2017} and the Dark Energy Survey \citep*[DES,][]{Flaugher2015} can now achieve levels of cosmological parameter constraints competitive with those from cosmic microwave background 
observations \citep{DES3x2_2019}. 
 
In order to reach such precision, the redshift distribution of the weak lensing source galaxies, and in particular 
the mean redshift of any tomographic redshift interval, must be very precisely constrained. In 
\citet*[][hereafter H18]{Hoyle2018}, we estimated that, in the four tomographic bins chosen for the weak lensing 
cosmology analysis with the first year of DES survey data (redshift binning $0.2<z<0.43$, $0.43<z<0.63$, $0.63<z<0.9$, and $0.9<z<1.3$), the mean redshifts are known to Gaussian uncertainties of 0.016, 0.013, 0.011, and 0.022 respectively. 
The anticipated scale of the full DES survey data implies that these uncertainties need to be reduced by roughly a factor of five, else they will overwhelm the statistical errors. Forthcoming experiments (LSST\footnote{\url{www.lsst.org}}, 
Euclid\footnote{\url{http://sci.esa.int/euclid/}}, and WFIRST\footnote{\url{https://wfirst.gsfc.nasa.gov/}}) require yet more stringent precision and accuracy.

In this context, a long literature has developed, describing approaches to derive redshift distributions 
\citep{Mandelbaum2008,Hildebrandt2012,Benjamin2013,Schmidt2013,Rau2015}, validate them 
\citep{Sanchez2014,Bonnett2016,Choi2016,Hildebrandt2018,Hoyle2018,Tanaka2018,Wright2018} and overcome 
some of the expected challenges in doing so \citep{Newman2008,Rau2017,Buchs2019,Sanchez2019}. 
A natural approach to validation is to use the very precise redshifts 
that can be obtained from spectroscopic observations of some of the science-sample galaxies, and almost all 
methods for deriving photometric redshift distributions are tested in this way. 
The principal challenges in validating with spectroscopy 
are mis-assigned spectroscopic redshifts \citep{Cunha2014, Newman2015}, colour and magnitude-dependent 
differences in sampling rate \citep{Lima2008} 
and sample variance arising from the fact that the spectroscopic objects tend to be located in small calibration fields -- often referred to as ``field-to-field variance'', or sometimes ``cosmic variance'' \citep{Cunha2012}. The first challenge is effectively solved by using only the highest confidence redshift determinations, transferring the problem to a greater imbalance in sampling rate.

\cite{Lima2008} presented an algorithm for estimating the redshift distribution of a target photometric sample from a spectroscopic data set directly, by accounting for the differences in sampling. It amounts to estimating the density of objects in the locale (in data space -- e.g. colour-magnitude space, including scatter due to photometric errors) of each spectroscopic data point, in both the spectroscopic sample itself and in the target data. The ratio of these densities is then given as a weight to the spectroscopic object. Finally, the redshift distribution is recovered by creating a weighted histogram of the spectroscopic sample. In addition to accounting for differences in sampling rate, weighting in this way also reduces the impact of field-to-field variance (e.g. H18). See also \cite{Sanchez2020} for a principled method to estimate and propagate the remaining sample variance in the resulting redshift distributions.

\begin{table}
\begin{center}
\caption{Characteristics of the main spectroscopic samples used in the DES Y1 analysis.}
\begin{tabular}{cccc}
\hline
Survey & Number of spectra &mean redshift &  total weight \\ \hline
VVDS & 11,121 & 0.60 & 0.15 \\
VIPERS & 9,455 & 0.58 & 0.13 \\
DEEP2 & 7,167 & 0.96 & 0.13 \\
zCOSMOS & 11,751 & 0.54 & 0.13\\
WiggleZ & 13,496 & 0.57 & 0.10 \\
3D-HST & 7,011 & 0.86 & 0.10 \\
ACES & 4,244 & 0.58 & 0.08 \\
OzDES & 12,436 & 0.61 & 0.06 \\
eBOSS$_{{\rm ELG}}$ & 4,322 & 0.96 & 0.03 \\
\hline
\end{tabular}
\label{tab:surveys}
\end{center}
\end{table}

In recent applications to weak lensing cosmology, this approach of weighting a spectroscopic sample has been used as either an independent 
measure of the redshift distribution \citep{Bonnett2016}, or as the leading redshift solution 
\citep[the so-called ``direct calibration'', or DIR method,][]{Hildebrandt2018,Joudaki2019,Asgari2019}. 
Insofar as a full and direct validation of photometric redshifts with spectroscopy goes, this weighting scheme is the 
only technique employed in weak lensing analyses to date.
One of the main requirements for using direct weighting of spectroscopic samples is that the spectroscopic data 
set must cover the same colour-magnitude space as the target sample (weak lensing sources, for example). 
However, there is the further assumption being made when taking such an approach, that within any given region 
of the colour-magnitude space the spectroscopic redshifts are representative of the local true redshift distribution of the lensing source sample. 
In general, there is no reason for this assumption to hold true. In any region of colour-magnitude space there will 
be some width to the redshift distribution --- perhaps due to photometric errors or insufficient dimensionality to the 
data. It is not difficult to conceive of situations in which low and high confidence redshift determinations in that region have
systematically different redshifts.

In \cite{Bonnett2016} we used existing spectroscopic and photometric datasets, similar to those considered in this paper, and showed that regions of colour-magnitude space that have poor spectroscopic success rates ($<65\%$) are on average biased by $\Delta z \sim 0.03$ with respect to the COSMOS photometric redshifts of a weak-lensing-like selection. This bias drops to $\Delta z \sim 0.01$ for regions with higher completeness. Similarly, work from \citet{Gruen2017} found that there exist significant biases (up to $\Delta z = 0.1$) in terms of the mean redshift, with the worst cases occurring at greatest depth.
In this work we take one step further and investigate the origin of this bias. We assess, via spectroscopic simulations, 
the magnitude of the bias in terms of the mean redshift of the inferred redshift distributions for a target photometric 
sample (designed to mimic a weak lensing survey). We use DES as an example but the principle can be applied to 
other experiments.

The paper is structured as follows. In \Sref{sec:sims} we describe the full procedure in constructing redshift distributions 
from a simulated spectroscopic sample. 
The target sample for which we wish to construct redshift distributions is the weak lensing sample for the first year of DES data 
(DES Y1). We describe the simulated spectra constructed from a full $N$-body simulation, the process of redshifting the spectra 
and determine the confidence level, enlarging the sample with a random forest approach, and ultimately reweighting the 
spectroscopic sample to match the photometry of the target sample. 
In \Sref{sec:results0} we present our findings in terms of the bias in the mean redshift in tomographic bins introduced via the 
incompleteness in the spectroscopic sample. We demonstrate first with VIPERS as an example of how targetting strategies introduce incompleteness, and then with a 
spectroscopic sample similar to that in DES Y1. In \Sref{sec:mitigation} we discuss three potential mitigation approaches of 
this bias and estimate their performances. We conclude in \Sref{sec:conclusion}.

\section{Reconstructing the redshift distributions via a simulated spectroscopic sample}
\label{sec:sims}

In this section, we simulate the full process involved in constructing the principal spectroscopic samples that overlap the 
DES Y1 footprint and that would be used to obtain redshift distributions in a DIR-like method. Below we first describe the 
spectroscopic data we aim to simulate (\Sref{sec:data}). Then we introduce the set of galaxy simulations that our work 
is built upon (\Sref{sec:buzzard}). Next we describe how we simulate the observed spectra (\Sref{sec:sim_spec}) as 
well as the process of having human ``redshifters'' visually inspect the simulated spectra and assign them quality flags (\Sref{sec:redshift}). 
Next, we train a random forest (RF) on these simulated spectra and assign quality flags to generate a larger dataset (\Sref{sec:RF}). 
The RF sample is then reweighted so that the photometry is matched to the target sample (\Sref{sec:reweighting}) where
we could compare the redshift distribution with the truth. Finally we discuss the various simplifications in this procedure and 
how they might affect our main results (\Sref{sec:simplifications}).

\subsection{Overview of spectroscopic data in DES}
\label{sec:data}

We use DES Y1 as an example survey to study in this paper, though the principle that we explore is applicable to any 
similar experiment. The DES Y1 weak lensing analysis carried out in 
\citet{DES3x2_2019,Troxel2018} used 26 million source galaxies to place unprecedented constraints on cosmological parameters. 
The basis of the redshift distribution of these source galaxies used in that analysis is detailed in H18. 

Table 1 and 
Appendix A in \citet{Bonnett2016} summarise the spectroscopic samples covered by the deep supernova fields in 
DES, which serve as calibration fields for the main-survey redshift distributions. As detailed in 
\citet{Sanchez2014} and \citet{Bonnett2016}, these photo-$z$ calibration fields were chosen to overlap with a 
number of key deep spectroscopic samples. In particular, three of the fields, SN-X1, SN-X3 and SN-C3, are well-studied 
extra-galactic fields containing VVDS Deep \citep{LeFevre2005}, ACES \citep{Cooper2012} and the rich spectroscopy built up in the SXDS / UKIDSS 
Ultra-Deep Survey (e.g. \citealt{Hartley2013}). The other two 
calibration fields were chosen to overlap with the VVDS Wide F14 field \citep{Garilli2008} and COSMOS \citep{Scoville2007}, 
which again provides a large number of spectroscopic samples for training. Overall, the dominant spectroscopic 
samples used in \citet{Bonnett2016} and H18 are: VVDS, VIPERS \citep{Guzzo2014}, DEEP2 \citep{Newman2013}, zCOSMOS\footnote{We use ``zCOSMOS'' 
to refer to the publicly available zCOSMOS-bright sample at $15<i<22.5$.} \citep{Lilly2009}, WiggleZ \citep{Parkinson2012}, 3D-HST \citep{Brammer2012}, ACES, 
OzDES \citep{Childress2017} and eBOSS \citep{Dawson2016}. \Tref{tab:surveys} lists the major characteristics of these samples. This table motivates the choices of spectroscopic 
samples we simulate later in this paper. That is, although these samples were not used as the primary redshift calibration 
method in H18, if DES Y1 were to use a direct redshift calibration method (or, the DNF algorithm, \citealt{DeVicente2016}), these would form the basis of the spectroscopic sample of choice.

In this paper, we choose to focus on VVDS, VIPERS and zCOSMOS. The data for these three 
spectroscopic samples were all taken by the same instrument, VIMOS \citep{LeFevre2003}, and are three of the four surveys 
that carry the greatest weight in our DIR-like algorithm.\footnote{Here, ``weight'' refers to the fraction of the photometric 
sample that is represented by galaxies in the survey after adjusting for representativeness, following \cite{Lima2008}.} 
VVDS is further split into ``Wide'' and ``Deep'' fields. 
We note that due to limitations in our simulations, we will not perform this study with the DEEP2 sample. The 
DEEP2 sample was taken via the DEIMOS spectrograph \citep{Faber2003}, and has the particular strength of high enough spectral resolution to split the [OII] doublet. The \texttt{k-correct} templates used for this work are at lower resolution, cannot replicate that strength and thus would not allow meaningful results to be obtained from our simulations.

\begin{table*}
\begin{center}
\caption{Observational parameters used to generate the simulated spectra.} 
\begin{tabular}{ccccc}
\hline
Survey & VVDS Wide & VVDS Deep &  VIPERS & zCOSMOS \\ \hline
Selection criteria & $17.5<i<22.5$ & $17.5<i<24$ & $17.5<i<22.5$;  & $15<i<22.5$ \\
                    &            &               &     $r-i>0.5(u-g)$ or $r-i>0.7$    & \\
Exposure time (sec) & 45  & 270 & 45 & 60 \\  
Number of spectra & 21550 & 12932 & 20452 & 20689  \\  
Sky emission model & \multicolumn{4}{c}{ESO VIMOS Exposure Time Calculator} \citet{Noll2013} \\  
Instrument transmission & \multicolumn{4}{c}{ESO VIMOS Exposure Time Calculator} \citet{LeFevre2003} \\  
Slit loss  & \multicolumn{4}{c}{30\%}  \\  
Mirror area (m$^{2}$) & \multicolumn{4}{c}{51.86} \\  
Star fraction & 0.0 & 0.0 &0.0488 & 0.0633\\  
\hline
\end{tabular}
\label{tab:obs_params}
\end{center}
\end{table*}

\subsection{Mock galaxy catalog}
\label{sec:buzzard}

The simulated spectroscopic surveys that we produce for our analysis are based on an initial selection from the 
Buzzard (v1.1) mock galaxy catalogue (\citealt{DeRose2019}, Wechsler et al. in prep). 
In this set of simulations, three flat $\Lambda$CDM dark-matter-only $N$-body simulations were used, 
with $1050^3$, $2600^3$ and $4000^3$ ${\rm Mpc^3 h^{-3}}$ boxes and $1400^3$, $2048^3$ and $2048^3$ particles, 
respectively. These boxes were run using \textsc{LGadget-2} \citep{Springel2005} with \textsc{2LPTic} initial 
conditions from \citet{Crocce2006} and \textsc{CAMB}. The cosmology assumed was $\Omega_{m}=0.286$, 
$\Omega_{b}=0.047$, $\sigma_{8}=0.82$, $h=0.7$, $n_{s}=0.96$, and $w=-1$. Particle lightcones were created 
from these boxes on the fly. Galaxies were then placed into the simulations and $grizY$ magnitudes and shapes assigned to each galaxy using the algorithm Adding Density Determined Galaxies to Lightcone Simulations 
(ADDGALS, Wechsler et al. in prep.). 
Each galaxy is assigned an SED from SDSS DR6 \citep{Cooper2006} by finding neighbours in the space of 
$M_{r}-\Sigma_{5}$, where $\Sigma_{5}$ is the projected distance to the fifth nearest neighbour in redshift slices of 
width $\Delta z = 0.02$. These SEDs are k-corrected and represented by five coefficients that correspond to 
five \texttt{k-correct} templates. The spectra are then integrated over the appropriate bandpasses to generate the 
DES $grizY$ photometric magnitudes. A further post-processing step is used to add appropriate photometric 
errors to the magnitudes according to what is measured in the DES Y1 data. 

\begin{figure*}
\begin{center}
\includegraphics[scale=0.65]{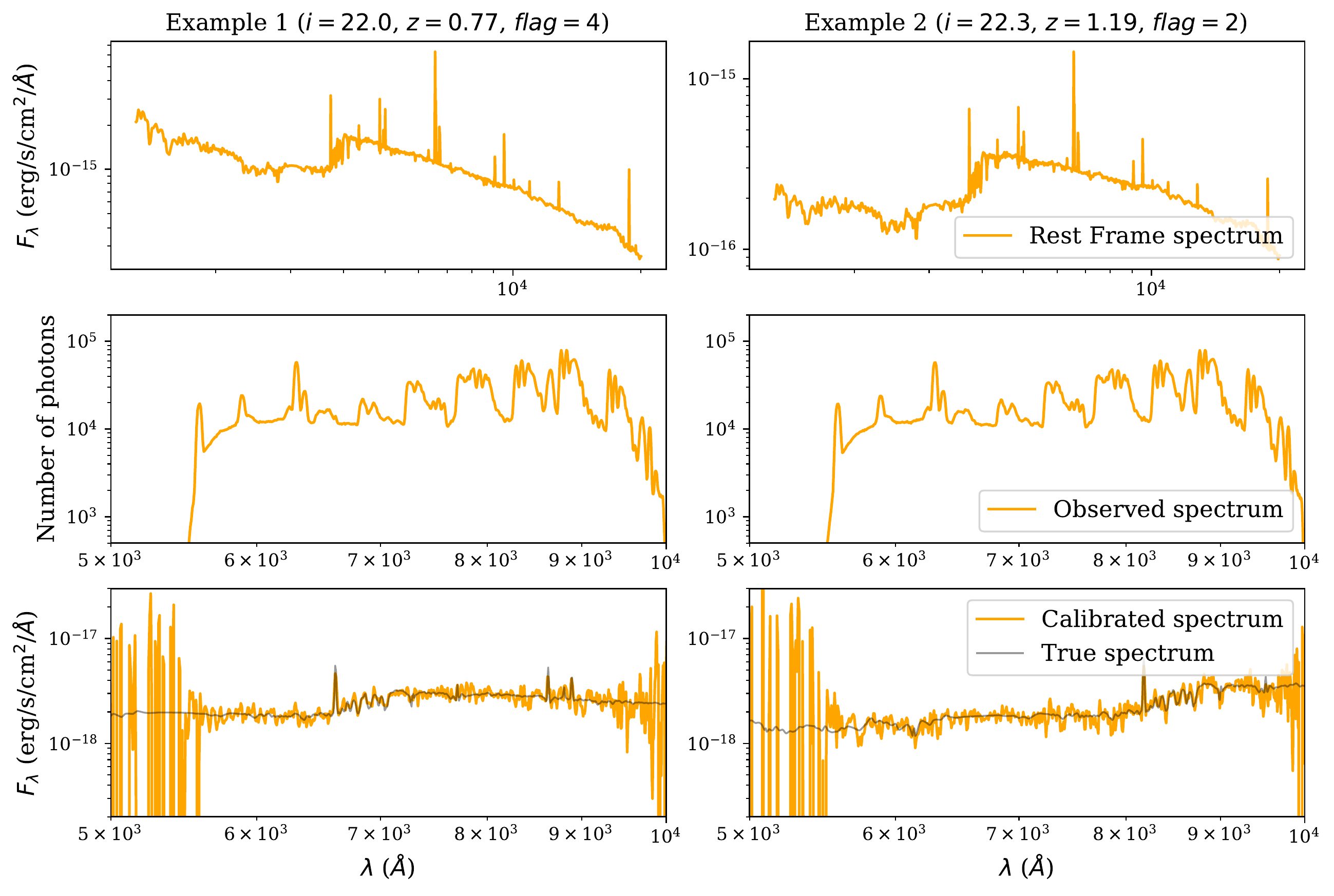}
\end{center}
\caption[]{Example spectra from the simulated VVDS Deep survey. {\bf Top panels:} Original rest-frame linear combination of 
\texttt{k-correct} components. {\bf Middle panels:} Poisson-sampled spectrum including sky emission. Note that the apparent shape 
of the spectrum is dominated by the sky emission. {\bf Bottom panels:} Final sky-subtracted and calibrated simulated spectrum 
(yellow) overlaid with the true spectrum (black). The former is what is passed onto the next stage for redshifting. 
The $i$-band magnitudes as well as the true redshifts for the two galaxies are listed at the top of the figure. The spectra on the left represents an example of a good spectrum (Flag=4), while the spectra on the right represent a spectrum of relatively 
poor quality (Flag=2). }
\label{fig:example_spec}
\end{figure*}

\subsection{Simulated spectroscopic training set}
\label{sec:sim_spec}

We now wish to construct simulated spectroscopic surveys from the mock galaxy catalogue described above.  
Specifically, this includes simulating the target selection function and the spectra of those targets with the expected 
signal-to-noise ratio.   

As described in \Sref{sec:data}, the four data sets of interest are: VVDS Deep, VVDS Wide, VIPERS and 
zCOSMOS. Simple magnitude and colour cuts form the target selection in each of these 
surveys (see \Tref{tab:obs_params} for the magnitude and colour cuts used in each sample). Our simulated 
data contains both noiseless, true photometry and simulated observed photometry, as described in \Sref{sec:buzzard}. 
Spectroscopic targeting was performed with the true magnitudes to reflect the fact that typically deeper data was used during 
the real target selection, and to avoid a one-to-one correspondence with the simulated DES photometry (again, to better 
reflect the real situation). The selections were sampled from areas of sky that are similar to the real survey data, and placed 
with similar angular separations as the real surveys, in order to capture the appropriate field-to-field variance uncertainties. 
However, note that this simple selection does not take into account other complexities in a real target selection scenario 
where e.g. slit mask constraints are an issue. 

Next, observational parameters of the different surveys need to be defined. We use the parameters listed in 
\Tref{tab:obs_params}. These are based on outputs from the ESO exposure time calculator for VIMOS, which uses sky illumination and transmission function defined in \citet{Noll2013} and \citet{LeFevre2003} respectively. We choose to apply average values for moon phase (grey) and slit losses ($30\%$) to all objects for simplicity.

To generate a simulated spectrum that mimics the noisy spectra of the real surveys after sky-subtraction and calibration, we start from 
the mock target list described above and carry through the following steps:
\begin{enumerate}
\item Multiply \texttt{k-correct} templates with coefficients provided in the mock galaxy catalogue to get a rest-frame spectrum.
\item Redshift the rest-frame spectrum to the galaxy's redshift $z$ so that the spectral flux density at $\lambda$ is shifted to $(1+z)\lambda$.
\item Re-bin the spectrum to the required instrument resolution.
\item Convert the flux density into units of photon counts per wavelength bin. 
\item Apply slit-loss factor and transmission.
\item Add sky background (which already includes the transmission efficiency).
\item Poisson sample the noisy spectrum (including sky).
\item Subtract an estimated sky background. In practice, the number of sky pixels in each slit means that this value is close to the true value.
\item Divide by the transmission and slit-loss factor to correct for instrument response and flux-calibrate the spectrum.
\end{enumerate}
This process is done for all the objects in all the surveys and packaged into FITS files that could be easily loaded into 
the redshifting program for the next step. See also \citet{Fagioli2018} for a similar process applied to simulate SDSS spectra.

In \Fref{fig:example_spec} we show two examples of simulated spectra 
from our simulated VVDS Wide data set. The top panels of \Fref{fig:example_spec} show the noiseless rest-frame
spectra. These example objects were chosen to be intrinsically fairly similar galaxies, but with different redshifts and confidence flags (explained in the next section). The middle panels 
show the ``observed'' spectra with noise and sky background. It is clear that the sky dominates the signal. The bottom 
panels show the calibrated spectra following corrections due to the transmission and estimated sky (yellow) and the 
true spectra (black). We see that the simulated spectra recover the shape of their respective input spectra very well in the range 5500-9900 ${\rm \AA}$. The galaxy in Example 1 is slightly brighter than that in Example 2, which results in a higher signal-to-noise ratio (S/N) spectrum.

We note that for practical reasons, we have made several simplifications in the above procedure (see a list of simplifications discussed in \Sref{sec:simplifications}). As a result, we expect the estimation from this analysis 
to be conservative -- further complications of the data should introduce higher redshift biases.

\subsection{Redshifting and Quality Flags}
\label{sec:redshift}

The spectroscopic surveys that we aimed to reproduce had redshifts determined by a combination of template cross correlation, emission line detection and a human inspection to confirm or replace those determinations. Importantly, the flags that represent the confidence that a given spectroscopic redshift is correct were all assigned by human observers. To be able to estimate the impact of any selection biases introduced, we must follow as close an approach as is practical and therefore also use human redshift and quality flag determinations.

Redshifting of the simulated spectra was performed by a team of eight observers with mixed levels of 
experience. Most were familiar either with redshifting optical spectra from AAOmega or VIMOS. Two of 
the eight had not performed such a task before and of our observers, two performed around 50\% of the redshifting. 
We use the software package \texttt{MARZ}\footnote{\url{https://samreay.github.io/Marz/}} \citep[Manual and 
Automatic Redshifting Software,][]{Hinton2016}, which is a web-based semi-automated template-fitting 
application, similar in essence to the commonly used \texttt{EZ} program for VIMOS \citep{Garilli2010}. \texttt{MARZ} 
uses a cross-correlation algorithm to match input spectra against a variety of stellar and galaxy templates 
in order to solve for the redshift. 

We generated a total of 75,623 simulated spectra (the sum of the number of spectra in all the samples in 
\Tref{tab:obs_params}), and randomly selected 12,000 for the redshifting procedure. These spectra were 
split into 40 subsamples, and each redshifter examined one or more of these subsamples and 
returned their results. We did not attempt to redshift all the spectra, as it was impractical to replicate the 
redshifting of three full VIMOS surveys, but instead choose to train a random forest using these 12,000 
spectra and generate the full dataset in \Sref{sec:RF}.  

The task for these redshifters is to assign a best redshift and an estimate of how secure they believe 
that redshift to be in the form of a redshift quality flag. Each flag value 
corresponds to a different confidence level of the determined redshift:
\begin{itemize}
\item Flag=4: essentially 100\% certain
\item Flag=3: 95 - 99 \% certain
\item Flag=2: 90 \% certain 
\item Flag=1: 50\% certain
\item Flag=0: a guess 
\end{itemize}
In addition, there is a special flag 6 (9 in the VVDS scale), which is for cases where there is a clear emission 
line, but insufficient supporting information to be able to tell which 
line it is -- i.e. there are a small number of possible redshifts, but the values can be quite different from one 
another. For most of this work we later re-assign these flags as 2.5 because they effectively sit in between 
flag 2 and 3 in confidence for the purposes of weak lensing experiments. In practice almost all flag 6 
objects were given the correct redshift, but because even a small fraction of wrong redshifts can cause biases, such objects are typically not used in analyses on real data.

In the spectra shown in \Fref{fig:example_spec}, Example 1 has been given the highest confidence flag. The strong and clear emission line is in a clean part of the spectral range, relatively unaffected by bright sky lines, and easily identified as [OII] based on the abundant supporting information: a break in the continuum, multiple Balmer absorption lines and weak but present [OIII]+H$\beta$ lines. In contrast, the ambiguity over whether the [OII] line is real or a sky subtraction residual in Example 2, together with the lack of convincing supporting evidence (lower S/N Balmer lines), results in a Flag=2 determination. Had the object been at significantly lower redshift a higher confidence would almost certainly have been assigned. This is a fairly common mode of failure in attempting to obtain a secure redshift. Other typical failure modes include: key emission lines in blue galaxies lying entirely outside of the spectral range, the $4000{\rm \AA}$ break of red galaxies lying outside of the range (which occurs at both high and very low redshift in the MR and LR-Red grisms) and dust extinction reducing the S/N of emission lines and/or the $4000{\rm \AA}$ break. Spectra of very low S/N can be problematic too, if they do not happen to have multiple clear emission lines within the spectral window. Finally, observers can occasionally find it difficult to assign confident redshifts if a spectrum presents unexpected features (e.g. those associated with the presence of an AGN), or if it is a blend of two objects at different redshifts. These factors result in a highly complex selection function in the joint space of galaxy type, redshift and luminosity.

The above procedure is clearly not objective. Different redshifters may have different scales in mind when 
assigning confidence flags to the spectra they inspect. In fact, depending on external environmental conditions, a single observer can change the confidence scale 
for the same object from one sitting to the next. In the real surveys that our simulations are based on, 
each spectrum was inspected by more than one observer and an arbitration procedure was followed in 
the case of conflicts in redshift and/or quality flags. While it cannot guarantee uniformity across a large 
data set, this process at least helps to reduce the subjective variance between observers. To achieve 
the same goal we introduced an overlap into the subsamples that are selected: we ensure that at least 
$10\%$ of each subsample is also present in another subsample. In this way, a fraction of the human-viewed 
spectra are redshifted by two or more observers, or even by the same observer twice or more.

Using these multiply-viewed spectra we standardise the observer flags in the following way:
\begin{itemize}
\item For each redshifter, we examine the internal consistency of their flags. That is, if they always 
gave consistent flags for the same object, they will have a higher rank.
\item This ranking for the redshifters will dictate which solution is accepted in the case of multiply-observed spectra.
\item In addition, each observer, in order of rank, is given a shift to all flag values equal to the mean of their difference 
with the highest rank observer that they share $>20$ objects with.
\item The result of this procedure is a set of new (decimal value) flags for each object, which extends beyond 4.
\end{itemize}
In \Sref{sec:results} we examine how the main results of this work change if we do not standardise the flags.

\begin{figure}
\begin{center}
\includegraphics[width=\columnwidth]{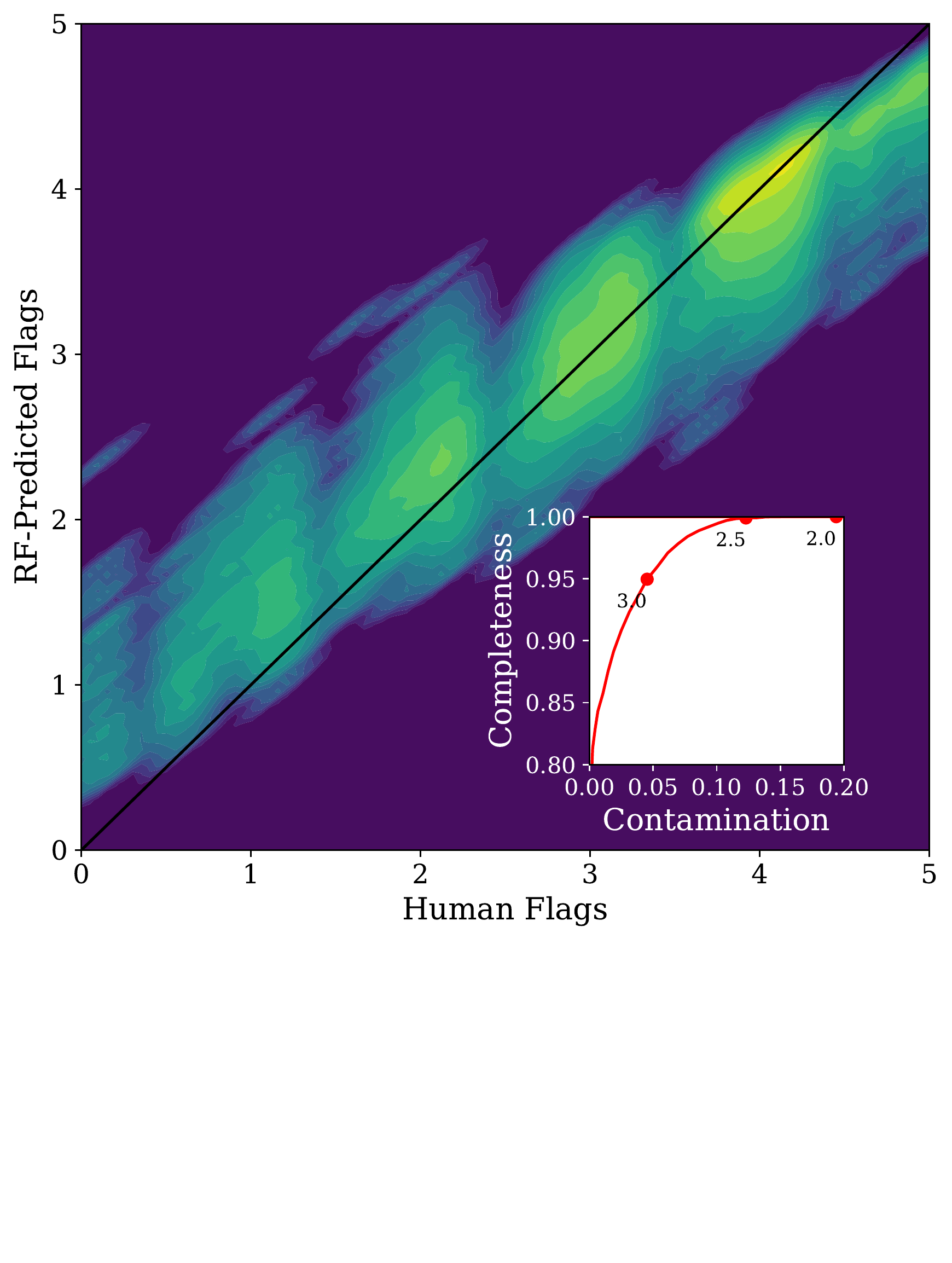} \vspace{-1.3in}
\end{center}
\caption[]{Random forest (RF) prediction of redshift quality flag against those determined by human observers. 
The mean predicted flags span a smaller range of values than the true flags, while the overall dispersion is 
of order 1. The bottom right inset shows a Receiver Operating Characteristic (ROC) curve of how well the RF performs in selecting objects to be retained or cut as the flag threshold value is changed. At the canonical threshold value of 3 the contamination by less secure objects and RF-induced loss of high-confidence objects are both fairly-well contained, at the $\sim5\%$ level. Samples cut with higher flag values are pure, but suffer a greater level of RF-induced incompleteness, resulting in a sample that is smaller than it should be. Conversely, at lower flag numbers the selected sample will be larger and more complete than it should be due to contamination by objects of intrinsically lower confidence. In \Sref{sec:results}, this ROC curve translates into a slightly over-estimated bias at the highest flag thresholds, and underestimated bias at lower flag thresholds.\label{fig:QOP_training}}
\end{figure}

\begin{figure}
\begin{center}
\includegraphics[width=0.85\columnwidth]{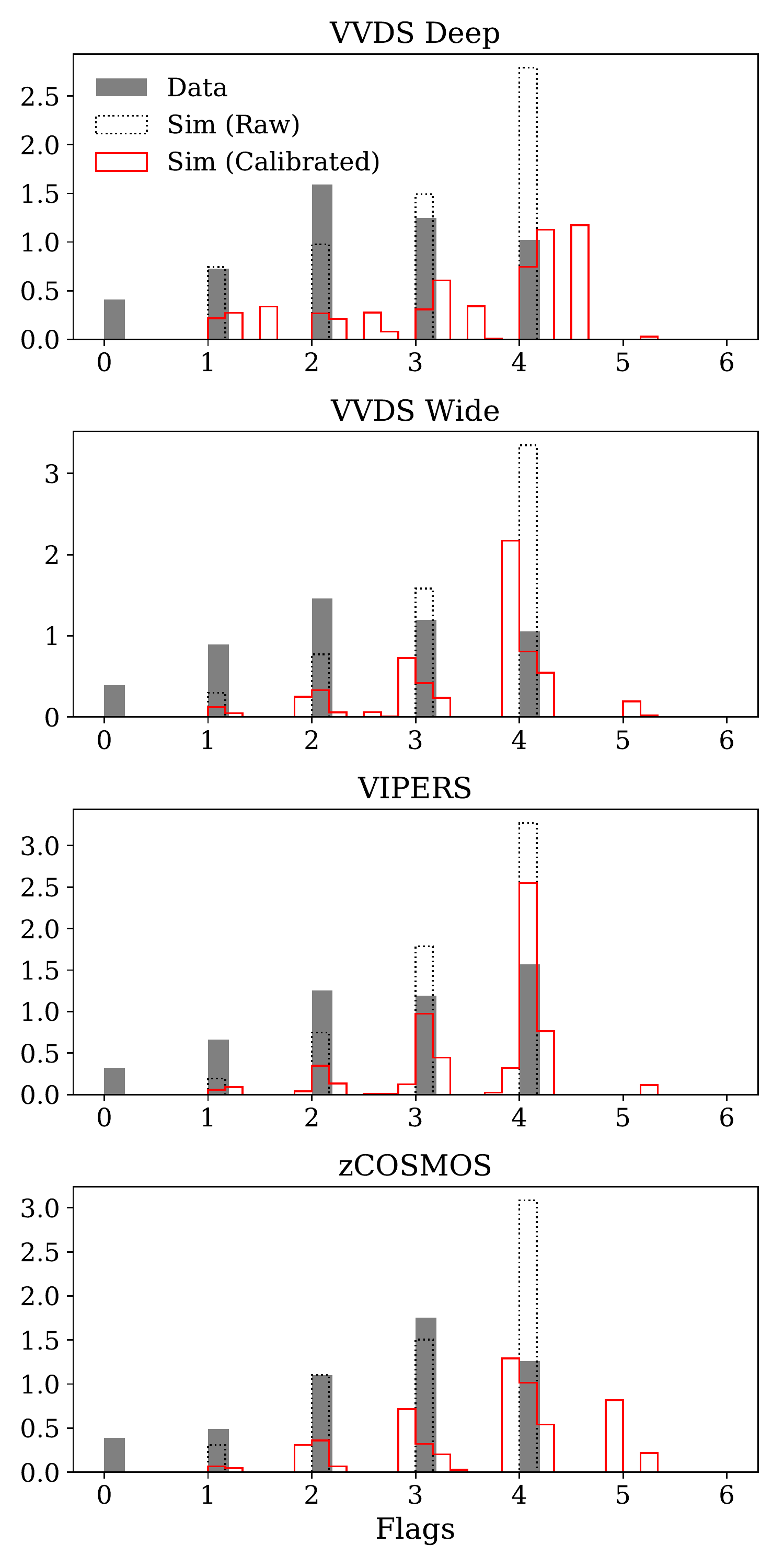}
\end{center}
\caption[]{Distribution of human-determined redshift quality flags for our simulated datasets (dotted), compared 
with those from the real survey data (filled grey). We also overlay the calibrated flags in red. 
\label{fig:QOP_hist}}
\end{figure}

\subsection{Generating the full spectroscopic sample through random forest}
\label{sec:RF}

We use the standardised set of redshifter flags (which we will refer to as ``human flags") obtained from the previous 
section as a training set to expand our sample to the full spectroscopic data set generated in \Sref{sec:sim_spec}. Specifically, we use a random forest (RF) in regression mode, with features computed from the simulated spectra and a single output (the redshift confidence flag).

For the training, the whole sample is used, irrespective of survey origin. The features that are used for the 
training are the S/N of emission and absorption lines, and the strength of the $4000{\rm \AA}$ break. In 
particular, we calculate the S/N of the spectra in the rest-frame wavelength window $\Delta\lambda=$100~${\rm \AA}$ 
around the absorption lines listed in \Tref{tab:features} in Appendix~\ref{sec:features}.

Our approach implicitly assumes that these are the features that humans use and ignores 
additional information such as shape of the continuum. In practice, continuum shape is sometimes used as 
supporting information in a redshift determination, but it is unlikely  
that it is sufficient to change a moderate confidence redshift into one of high confidence. 

In \Fref{fig:QOP_training}, we show RF flags for the sample where human flags are also available. 
We note that they are strongly correlated, but with a scatter width $\sim1$ and a mean slope of 
the distribution that is smaller than 1. The inset panel shows a partial Receiver Operating Characteristic (ROC) 
curve of the RF-predicted flags, showing the completeness vs. contamination of the RF sample for different 
threshold values of the predicted flag (Flag=2, 2.5, 3 is marked on the curve). In \Sref{sec:results} we show that our results are largely insensitive to whether we use human or RF flags.

In \Fref{fig:QOP_hist} we compare the distribution of the quality flags from our simulated dataset with real distributions from VVDS, zCOSMOS and VIPERS. It is clear that our determinations are on average of higher confidence class. This is 
due to: 1) the fact that the simulated spectra are somewhat idealised and 2) differences between the confidence indicated by a given class between surveys -- for instance, our Flag=3 corresponds more closely to Flag=2 in VIPERS. In the later analyses we investigate the redshift bias as a function of the flag threshold -- one could adopt a higher flag threshold to account for the idealistic aspects of the simulations and any variance in the meaning of the confidence flags.

\subsection{Re-weighting and the target sample}
\label{sec:reweighting}

As we mentioned in \Sref{sec:intro}, when using spectroscopic redshifts directly to infer the redshift distribution in weak 
lensing surveys, common practice is to re-weight the sample, following e.g. \citet{Lima2008}, to account for any mismatch 
in the distributions of photometry between the spectroscopic and weak lensing data sets. We implement 
such a procedure, re-weighting the spectroscopic sample in our simulations (constructed by applying a given quality flag cut 
on the spectroscopic sample described in previous sections) to a target weak lensing sample. Weighting is performed with a k-nearest neighbours algorithm in four-dimensional colour-magnitude space ($g-r$, $r-i$, $i-z$, $i$-band magnitude), reflecting the DES survey observing strategy.

Throughout the paper, we assume four target samples -- matching the four tomographic weak lensing source samples 
used in H18. We assign all the galaxies to four tomographic redshift bins ($0.2<z<0.43$, $0.43<z<0.63$, $0.63<z<0.9$, 
$0.9<z<1.3$) via the mean redshift of the $p(z)$ output by the Bayesian Photometric Redshifts code \citep[BPZ,][]{Benitez2000}, 
which is run on the simulated `observed' fluxes. The BPZ 
set-up and binning scheme above follows the one used for the DES Y1 cosmology analysis H18. An $i<23.4$ cut is also 
applied to all the four samples -- this roughly mimics the weak lensing source galaxy selection, and is the target sample 
used throughout the paper, though real source catalogues have a softer magnitude cut due to the complexities of morphology and brightness in the selection cuts. In \Tref{tab:target_sample} we list the characteristics of this target sample compared to the DES 
Y1 weak lensing sample. Our target sample is slightly fainter on average than the DES Y1 sample. However, there is a tail that extends to fainter magnitudes in DES Y1 that our target sample does not include. These two contrasting differences mean that our target sample is fairly consistent with the DES Y1 sample in the mean redshift.
Also note that the spectroscopic sample described previously is selected from a sub-region of this target sample.

\begin{table*}
\begin{center}
\caption{Characteristics of the weak lensing sample in DES Y1 compared to that used in this paper with a simple magnitude cut $i<23.4$. 
In this table the first number is for the DES Y1 sample and the second number the target sample used in this paper.}
\begin{tabular}{ccccc}
\hline
Survey & $0.2<z<0.43$ & $0.43<z<0.63$ & $0.63<z<0.9$ & $0.9<z<1.3$ \\ \hline
Mean magnitude &  21.4; 21.7 & 21.7; 22.1 & 22.0; 22.3 & 22.5; 22.8 \\  
Mean redshift  & 0.38; 0.35 & 0.51; 0.45 & 0.74; 0.74 &  0.96; 0.99\\  
\hline
\end{tabular}
\label{tab:target_sample}
\end{center}
\end{table*}

\subsection{Simplifications in our approach}
\label{sec:simplifications}

Our simulations and analysis approach are idealised in several aspects. We discuss the simplifications in this section, 
but note that the purpose of this study is not to simulate a high-fidelity spectroscopic sample and estimate the exact 
value of the bias due to spectroscopic incompleteness (nor is it practical to do so). 
Rather, we use reasonable assumptions to illustrate the point that spectroscopic samples used to calibrate 
weak lensing surveys can in principle be biased due to selection effects in constructing the spectroscopic sample 
itself, even after re-weighting is applied. With the simulations we can also estimate the order of magnitude of this 
effect and compare with other systematic uncertainties in the redshift distributions. We also note that the simplification 
of the simulated spectra generally leads to a conservative estimate of the bias (i.e. the true bias in the data is likely to be higher).  

The first class of simplifications are those associated with simulating the spectra:
\begin{itemize}
\item We do not include the particularly severe red fringing that is seen in early VIMOS surveys. 
\item We assume fixed sky spectra and perfect knowledge of the transmission curve.
\item We ignore instrument flexure, mis-aligned slit masks and poor flux calibration. 
\item The spectra are based on \texttt{k-correct} templates, which only produce a limited range of unique spectra.
\item The Buzzard simulations themselves do not include all galaxy types as they are matched to a limited population 
and redshift of galaxies in SDSS.
\item We specify each survey with only the parameters listed in \Tref{tab:obs_params}.
\end{itemize}

The second set of simplifications and approximations are those made in the process of generating the quality flags for 
the full spectroscopic sample:
\begin{itemize}
\item We use the RF flags, which differ slightly from the flags that would be determined by a human redshifter.
\item We account for the observer - observer scatter by the simple priority and standardisation scheme described in \Sref{sec:redshift}
\end{itemize}

Finally, to cleanly isolate the effect of the spectroscopic selection effects from the photometric redshift estimation algorithm:
\begin{itemize}
\item We use the true redshift instead of the estimated redshift when evaluating the bias in the mean redshift. Implicitly this choice also avoids the need to simulated complex survey-dependent effects such as blending, which can result in an incorrect, but confident, redshift assignment based on the wrong spectral features (e.g. \citealt{Masters2019}).
\end{itemize}

Some of the effects we neglect, e.g. poorly aligned slit masks, will produce redshift failures that are essentially random as to which objects they impact. In such cases our estimated redshift biases would not change, simply the level of shot noise in the analysis would increase if we were able to model those effects correctly. Other simplifications will have the effect of making the human redshifter's job easier, and hence result in a more confident flag assignment. By using the canonical flag threshold of Flag $\ge3$ to determine the high-confidence sample (where appropriate) we are therefore being over-inclusive as to which objects are retained and thus under-estimating the magnitude of any redshift bias.

\section{Quantifying Spectroscopic Incompleteness Bias}
\label{sec:results0}

\subsection{VIPERS: survey-constructed incompleteness}
\label{sec:vipers}

Before examining the impact of incompleteness in our simulated data, we briefly revisit sample selection effects in 
spectroscopic surveys. In \cite{Bonnett2016} we showed that the artificial upper redshift limit that was used for 
determining spectroscopic redshift solutions in the PRIMUS dataset \citep{Cool2013} is propagated by machine 
learning algorithms to the redshift distribution estimation  
of the science sample. As a result, PRIMUS 
redshifts were not used in determining the redshift distributions of the weak lensing samples in \cite{DES2016}. 
Because of the size of the PRIMUS sample (88,040 galaxies) and the fact that this bias was imposed during redshift 
determination, the effect was rather clear and could not be compensated for by applying weights 
(e.g. \citealt{Lima2008}). However, spectroscopic samples are frequently selected for specific science purposes and 
many of the remaining samples contain biases of their own, for instance due to using colour cuts to isolate particular 
redshift intervals. If an employed colour cut is not available to a particular weak lensing experiment, then it is possible 
for small biases in redshift to be introduced during re-weighting, purely due to the projection in colour space. We 
demonstrate this issue using the simulated VIPERS spectroscopic sample described in \Sref{sec:sim_spec} as an example. 
We will also for illustration purpose use a target sample that is different from what we use in the main analysis.

The VIPERS team used a pair of selection criteria in colour space to broadly separate objects at $z>0.5$ from the 
lower-redshift population based on an initial $i$-band-selected catalogue. Identifying objects in this way enabled a 
very efficient survey strategy, due to strong spectral features falling within the spectral window of the LRred grism 
on VIMOS. The final dataset is large with high completeness ($90.6\%$ at redshift confidence $>96\%$\footnote{This estimate 
includes slightly lower confidence flags than typically used for weak lensing analyses.}, \citealt{Guzzo2017}) 
with just 45 minutes of exposure time per target. In the DES final redshift catalogue \citep{Gschwend2018}, 
there are similar numbers of objects from the VIPERS dataset and from the pure $i$-band selected sample of VVDS wide ($17.5<i<22.5$).

\begin{figure}
\begin{center}
\includegraphics[width=0.9\columnwidth]{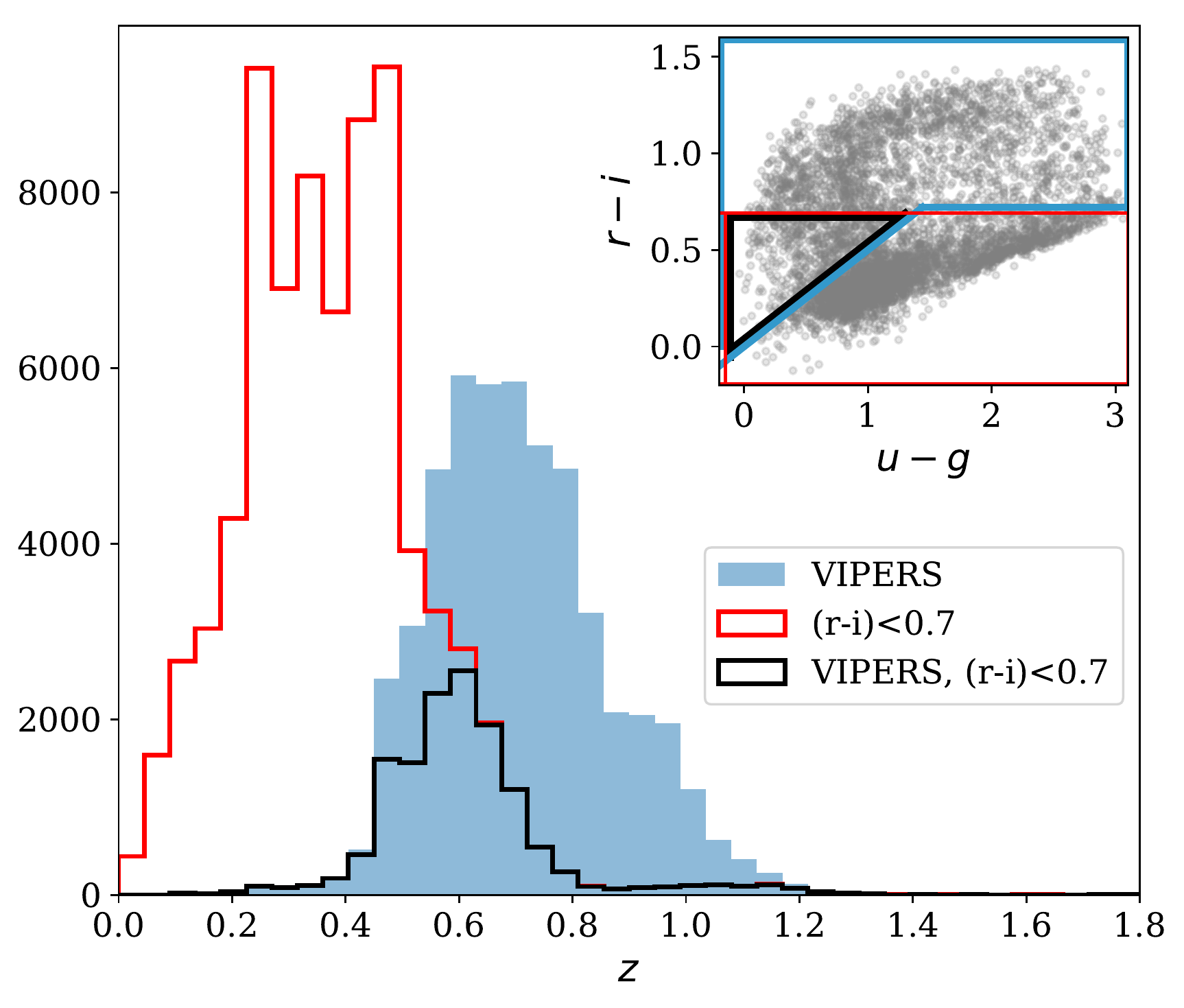}
\end{center}
\caption[]{Redshift distribution of galaxies matching the VIPERS colour selection: $(r-i)>0.5(u-g)$ or $(r-i)>0.7$ (solid), 
an $(r-i)<0.7$ sample and a sample selecting just where VIPERS overlaps at $(r-i)<0.7$. These latter two samples have 
different redshift distributions, and so re-weighting without $(u-g)$ colour information will result in biases. {\bf Inset:} these 
three samples in $(u-g)$ vs. $(r-i)$ colour space. \label{fig:VIPERS}}
\end{figure}

To illustrate the issue, we first apply the i-band selection criterion, $17.5<i<22.5$, to the Buzzard galaxy catalog. 
This sample will be used as the target sample here, note that the selection criteria is identical to that of VVDS wide 
(see \Tref{tab:obs_params}). The distribution of this sample in $(r-i)$ vs. $(u-g)$ colour space is shown in the inset of 
\Fref{fig:VIPERS}, together with the VIPERS selection criteria (blue outline, $(r-i)>0.5(u-g)$ or $(r-i)>0.7$) 
and two other colour-defined subsamples (black and red outlines). The redshift 
distributions of the galaxies in these three samples are shown in the main panel of \Fref{fig:VIPERS}. As expected, the 
vast majority of $z>0.5$ galaxies in the red selection region are contained within the black selection box (and hence 
within VIPERS), and the black region contains almost zero low-redshift galaxies. We assign all the galaxies to four 
tomographic redshift bins as described in \Sref{sec:reweighting}.

\begin{figure}
\begin{center}
\includegraphics[width=\columnwidth]{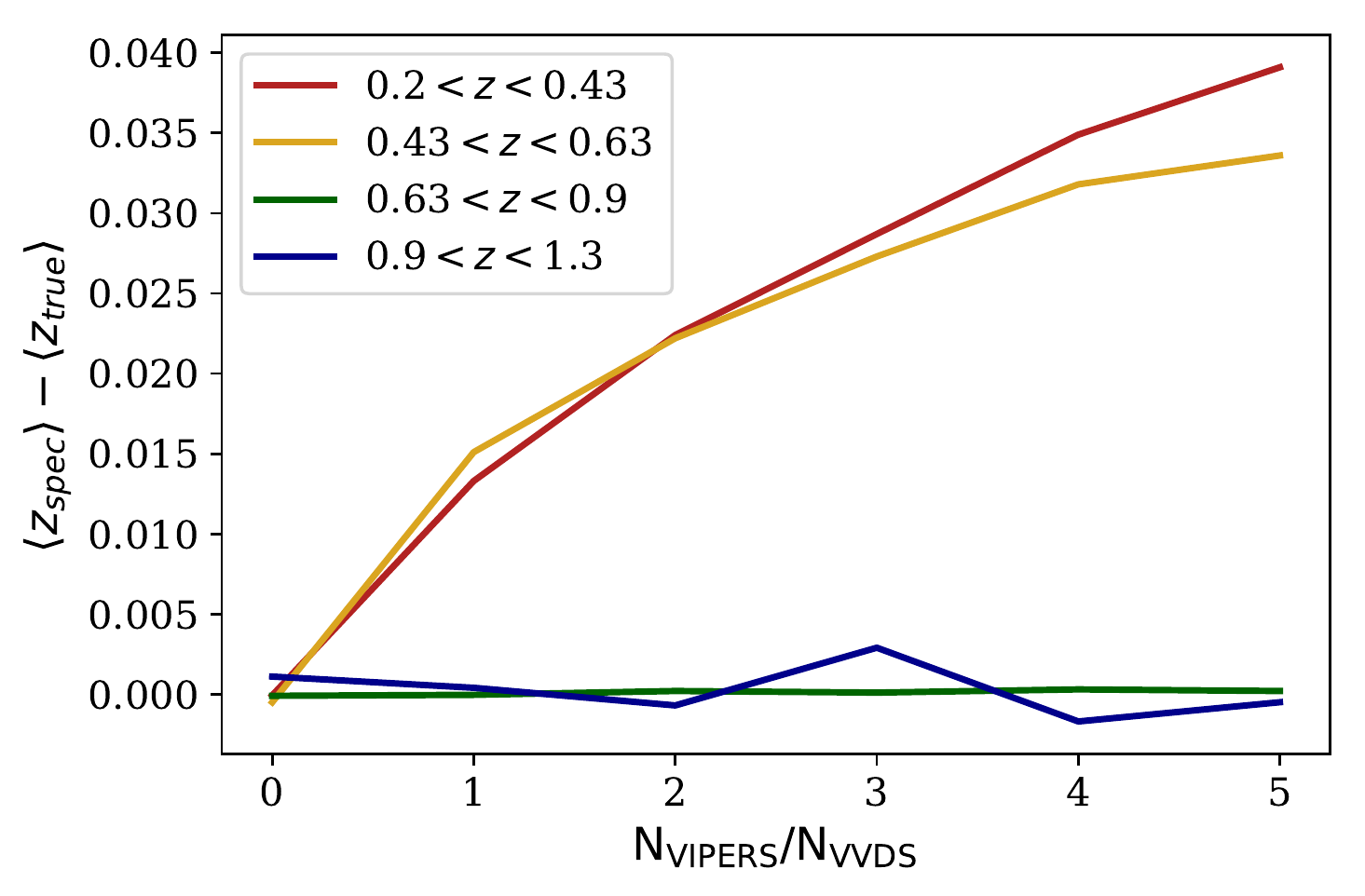}
\end{center}
\caption{Bias in the mean of the redshift distribution for four tomographic bins between 
a galaxy sample consisting a mix of VIPERS and VVDS wide galaxies and our target sample selected through a 
simple $17.5<i<22.5$ selection. A weighting scheme is applied to the redshift distribution of the galaxy sample to 
account for the difference in the colour-magnitude distribution in the VIPERS/VVDS sample and the target sample. 
From left to right, we vary the relative fraction of VIPERS and VVDS galaxies. 
The mean redshift is biased high in the two lower redshift bins when a significant fraction of the sample comes 
from VIPERS.\label{fig:bias_table4}}
\end{figure}

Next, we calculate the redshift bias for a sample of galaxies that contain a mix of VIPERS and VVDS wide galaxies. 
The redshift bias is defined as the difference in mean redshift of the weighted spectroscopic sample and the target 
sample. The weighting accounts for the difference in the colour-magnitude distribution in the sample compared to the 
target sample, but only in the colour-magnitude space that is available to DES photometry ($griz$). 
Negative redshift bias values indicate the spectroscopic sample is biased towards low redshift. \Fref{fig:bias_table4} 
gives the bias in each tomographic bin where the target sample is always the full sample with the $17.5<i<22.5$ 
magnitude selection. The x-axis corresponds to different ratios in the number of VIPERS sources to VVDS wide sources. 
The samples are assumed 100\% complete and occupy the same region of sky -- hence the bias is due purely to the 
imprint of the VIPERS colour selection function. 

We note here that when only VVDS wide galaxies are used 
($N_{\rm VIPERS}/N_{\rm VVDS} =0$), the spectroscopic sample is just a subset of the target sample, therefore the reweighting is perfect and the bias is essentially zero.

However, as we move to the right on the x-axis in \Fref{fig:bias_table4}, we see that the two low redshift bins become 
more biased as the fraction of VIPERS galaxies increase in the sample -- the bias is around a couple of per cent and 
increases the mean redshift of the bin. This is due to the lack of low redshift VIPERS galaxies that did not get properly 
compensated with the weighting scheme. On the other hand, the biases in the high redshift bins are much smaller and 
consistent with being simple noise fluctuations. In DES Y1 data, $N_{VIPERS}/N_{VVDS}\sim1$, suggesting a $\sim$0.015 
bias in the mean redshift in the low redshift bins.
One way to think of this selection function is that relative to a complete $17.5<i<22.5$ sample, the spectroscopic 
data set is systematically incomplete at red $u-g$ colours and blue $r-i$ colours. The crucial point here is that 
information is used to select objects that is not accessible to the survey, and hence the selection cannot be 
compensated for.

There is similar potential for a selection-induced bias in relying heavily on the DEEP2 data set. The DEEP2 team used a set of softened colour cuts to pre-select galaxies at $z>0.75$ in three of their four fields, including the one covered by DES data. The one control field with a purely magnitude-limited selection is in the extended Groth strip, and therefore too far North for DES, KiDS or LSST. The DEEP2 colour cuts used B, R and I-band data, and therefore cannot be reproduced in the DES photometry. We might anticipate a smaller bias than shown for VIPERS, on account of the fact that the missing band (B-band) is closer in effective wavelength the bluest DES band (the g-band) than in the case of VIPERS, where it is the U-band that is missing. However, it will also depend on the relative number of spectroscopic objects used and thus a reliable estimate would require that a realisation of the selection strategy be performed. 

We can draw a direct analogy from the VIPERS example above to generic selections of spectroscopic samples -- it 
is the strength of the available spectral features that determine whether a galaxy can be used for redshift validation. 
These spectral features are similarly information used for the selection that 
cannot be accessed with the photometric survey data. We will now examine the impact of making selections 
on spectroscopic information.

\subsection{Impact of incompleteness}
\label{sec:results}

Having examined the simple case in \Sref{sec:vipers}, we now move on to the full spectroscopic sample constructed in 
\Sref{sec:sims}. We combine the four simulated surveys as is typically done with real data: naively concatenating the data 
sets, cutting galaxies below some redshift flag criterion and then giving each object a weight such that the overall 
sample mimics the colour-magnitude distribution of the target sample (here, we weight in $griz$, as done in DES Y1). 
We then compute the difference in mean redshift between this sample (constructed to be incomplete due to the imposed flag cut) and the target sample (four $i<23.4$ 
magnitude-limited tomographic bins at $0.2<z<0.43, 0.43<z<0.63,  0.63<z<0.9, 0.9<z<1.3$), using the true redshifts of 
the simulated galaxies -- i.e. we assume 
that human redshift determinations are infallible. In our data this was indeed the case for the higher quality redshifts (Flag $\ge 3$), but is in general not the case, even for the highest confidence objects. For instance, blends of multiple objects at different redshifts lead to ambiguities, or even assiging a fairly bright galaxy the redshift of a much fainter one due to only one of the galaxies exhibiting emission lines \citep{Masters2019}. The spectroscopic sample is re-weighted as described in \Sref{sec:reweighting} and an estimate of the redshift 
distribution for the target sample is derived.

\begin{figure}
\begin{center}
\includegraphics[width=\columnwidth]{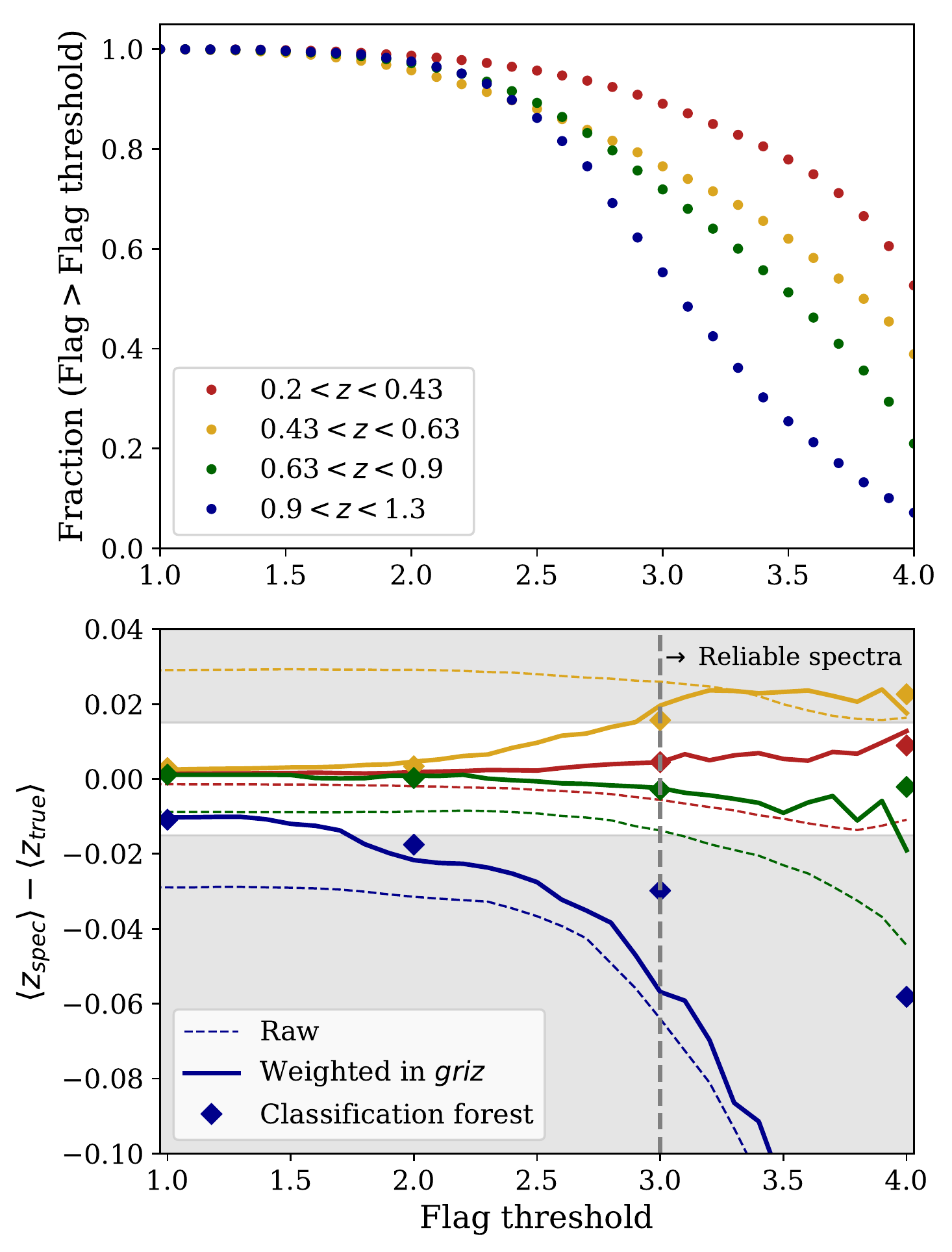}
\end{center}
\caption[]{{\bf Upper panel:} Survey completeness as a function of flag threshold in each tomographic redshift interval, 
split using the mean redshift derived by BPZ, as performed in the main DES analysis. {\bf Lower panel:} Bias in 
mean weighted redshift as a function of flag threshold, using the galaxies' true redshifts. The spectroscopic sample
is weighted to match the colour-magnitude space of a sample complete to $i<23.4$ in the same simulation fields.
The bias in the mean redshift comes from systematic incompleteness in the spectroscopic sample. The diamond markers 
show the same results using unstandarised flags. We also highlight the 
region $|\Delta z|< 0.015$, which is the approximate uncertainty in the DES Y1 photo-z from COSMOS15 calibration. 
\label{fig:bias_main}}
\end{figure}

\begin{figure}
\begin{center}
\includegraphics[width=0.95\columnwidth]{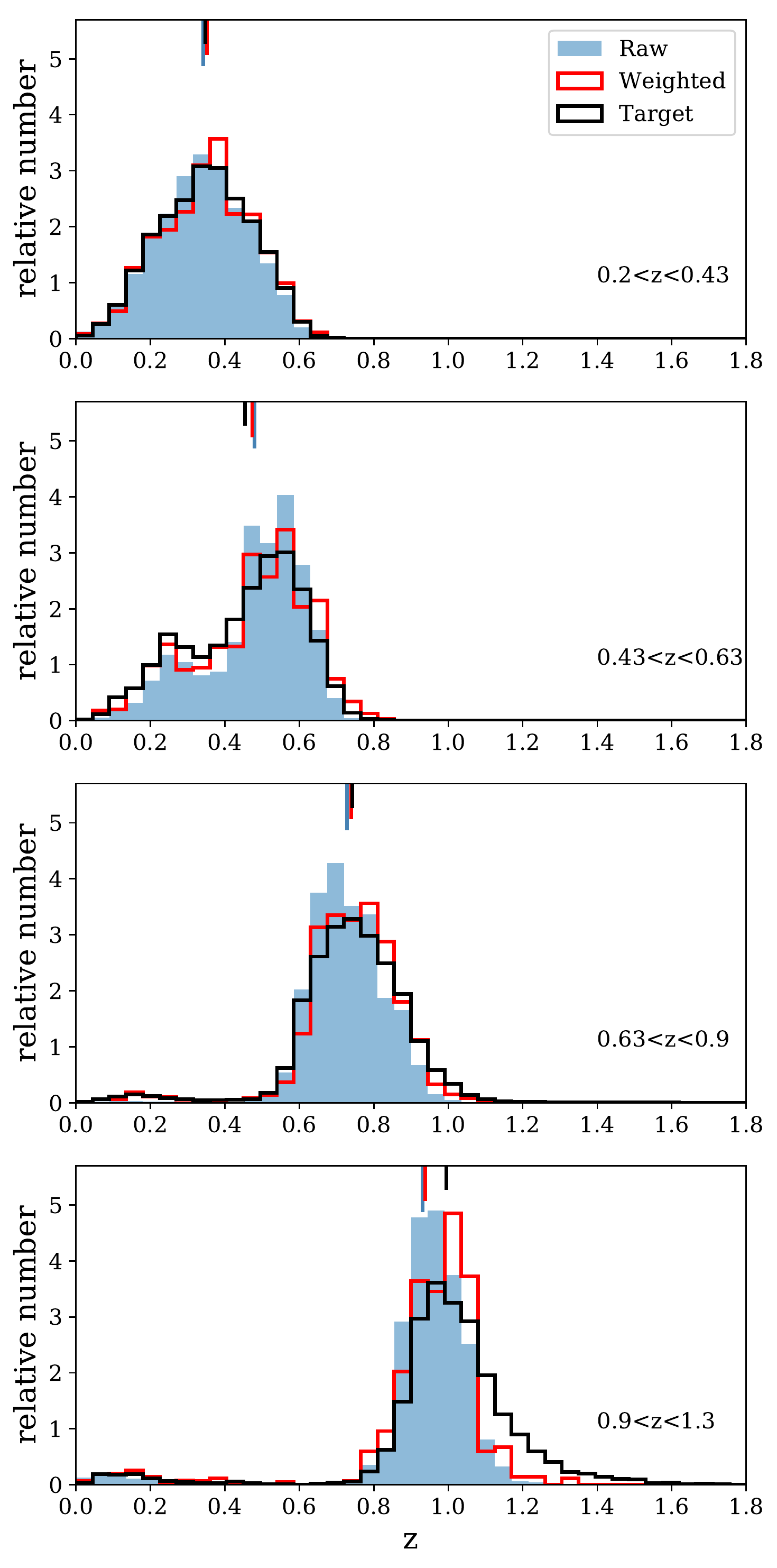}
\end{center}
\caption[]{Redshift histograms for complete (black), incomplete at Flag $< 3$ (blue shaded) and weighted (red) 
samples. The weighting matches the incomplete sample to the complete one in colour-magnitude space, but 
results in residual mismatches in the redshift distribution. The bars on the top of each panel mark the mean 
redshift for each of the distributions. 
\label{fig:flag3_hist}}
\end{figure}

\Fref{fig:bias_main} shows the completeness and $\langle z_{\rm spec} \rangle - \langle z_{\rm true} \rangle$ as a function of the flag threshold. 
Here $\langle z_{\rm spec} \rangle$ is the mean redshift for the unweighted and weighted spectroscopic sample after some flag threshold 
selection and $\langle z_{\rm true} \rangle$ is the mean redshift of the target sample.
We use a simple $i<23.4$ target sample and the tomographic redshift bins as described in \Sref{sec:reweighting}. The target 
sample in \Fref{fig:bias_main} covers the same fields as the spectroscopic survey. This is not to say that it is entirely free from 
sample variance effects, and in fact $\langle z_{\rm spec} \rangle - \langle z_{\rm true} \rangle$ is nonzero even when 
objects with very poor flags are included in the spectroscopic 
sample (Flag $\sim 1$). After weighting, however, $\langle z_{\rm spec} \rangle - \langle z_{\rm true} \rangle$ is reduced to $< 0.01$ across all tomographic bins. Raising the 
flag threshold we begin to see a degradation in mean redshift recovery with respect to the complete sample case, even after weighting. 
At a nominal high-confidence cut, Flag $\ge 3$, the bias is still small ($< 0.01-0.02$) in the three lower-redshift bins; though we 
note that this level of error is already barely within the target accuracy. The highest redshift bin suffers a substantial bias of $\sim 0.05$.
An error of this size means that incomplete spectroscopic samples such as those considered here are a poor choice for 
validating high-redshift samples. We show in \Fref{fig:flag3_hist} redshift histograms for the samples cut at confidence Flag $\ge 3$. 
Visually, we see that the reweighting scheme performs well in recovering the shape of the redshift distributions in all bins, 
correcting differences in the tails as well as the shape of the core distribution. However, there are some small differences 
remaining, resulting in the overall error in the mean redshift. 

The fact that the effect of spectroscopic incompleteness is most severe in the high redshift tomographic interval is hardly a surprise. The range 
$0.9<z<1.3$ includes galaxies with important strong spectral features -- the $4000{\rm \AA}$ break and [OII] emission line 
doublet -- buried in bright sky lines or even falling outside the useful spectral window of the red VIMOS grisms. Galaxies at 
these redshifts are also fainter on average than those in the lower redshift subsamples, and the increased photometric errors act 
to broaden the redshift distribution at any given location in colour space. The combination of redshift-dependent incompleteness 
at fixed colour and reduced ability to localise objects in colour-magnitude space results in weighting being ineffective. 

As noted above, the y-axis in the bottom panel of \Fref{fig:bias_main} (lower panel) contains both the effect from the flag threshold cut and 
the fact that there is field-to-field variance between the four spectroscopic surveys we are simulating. In Appendix~\ref{sec:vvdsdeep} 
we isolate the two effects and show that our main conclusions remain the same.

We have also tested the full analysis using unstandarised (integer) flags and show the results as diamond markers in 
\Fref{fig:bias_main} and \Fref{fig:bias_vvdsdeep}. Overall the predicted biases are lower with unstandarised flags and look visually 
more similar to the human flags described later in \Sref{sec:depth}. However, we decide to use the RF flag in our fiducial analysis 
because 1) the flags in real data are not always integers and 2) the flag assignment (and therefore the exact structure 
in the curves in \Fref{fig:bias_main}) is arbitrary and survey-dependent. In addition, since the bias in the highest redshift bin with 
the unstandarised flags still exceeds the requirement, our main conclusion does not change qualitatively. Essentially we can view 
the standarised and unstandarised flags to bracket the range of biases we expect.

\begin{figure*}
\begin{center}
\includegraphics[width=\columnwidth]{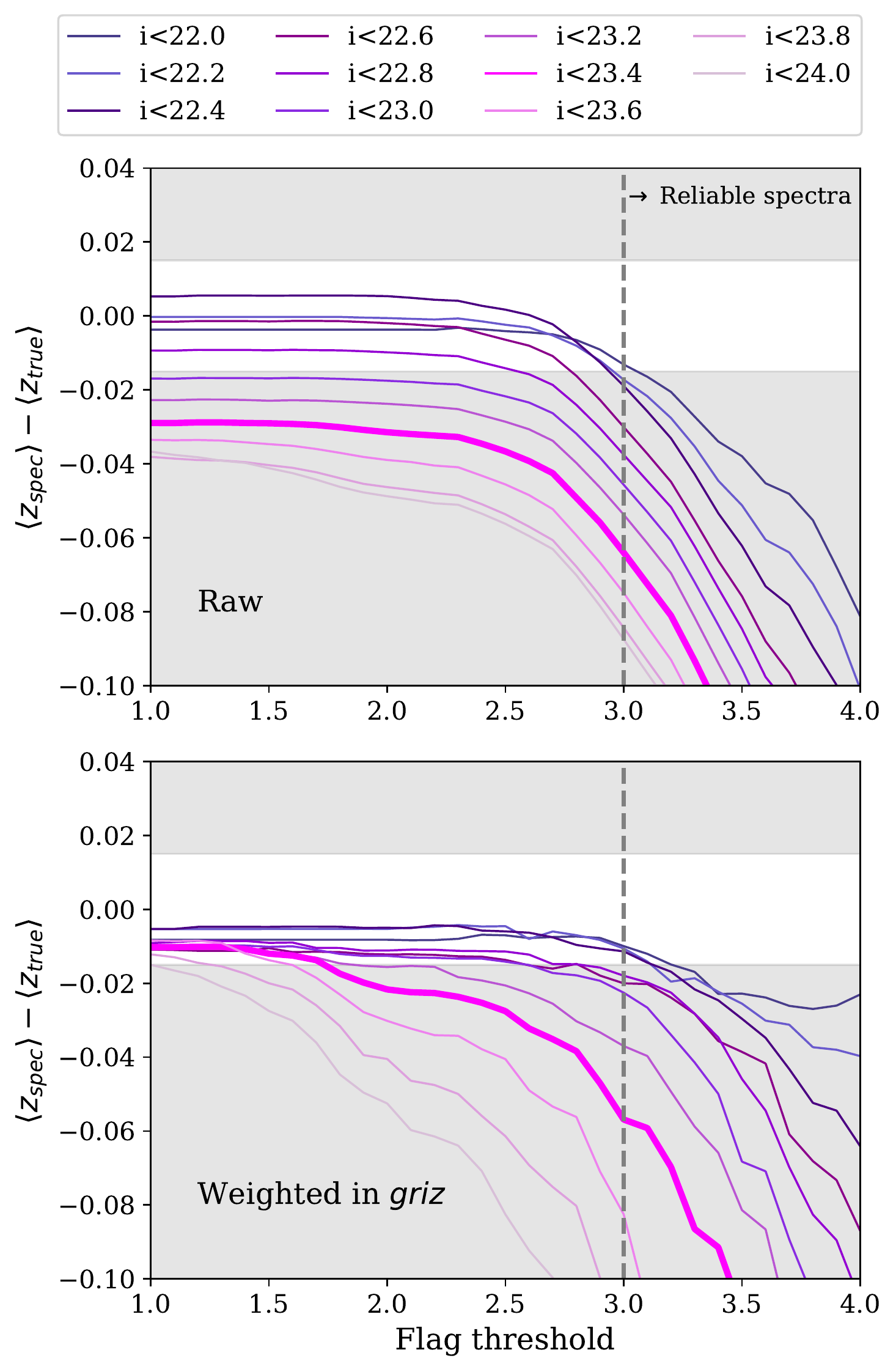}
\includegraphics[width=\columnwidth]{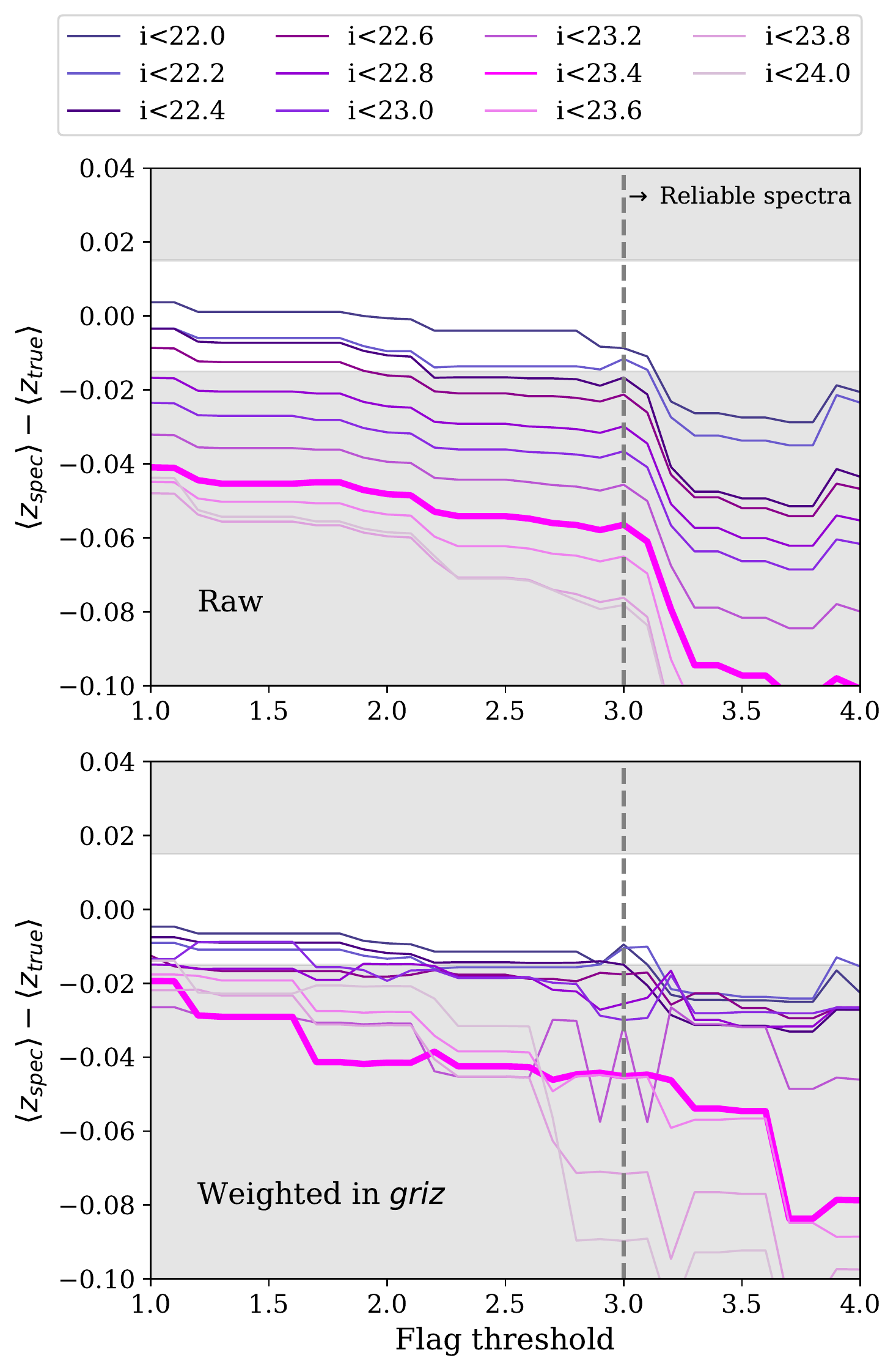}
\end{center}
\caption[]{\textbf{Left panels:} Redshift bias from spectroscopic incompleteness as a function of flag thresholds for target samples 
of different depth, at $0.9<z<1.3$. The upper panel shows the bias from the raw sample while the lower panel shows 
the bias in the reweighted sample. We highlight the region $|\Delta z| < 0.015$, which is the approximate uncertainty in the DES Y1 
photo-$z$ from COSMOS15 calibration. We also mark flag threshold = 3, which is the typical flag  
adopted to construct spectroscopic samples. \textbf{Right panels:} Same as the left panel but only using human redshift 
flags and not RF flags. In both panels, the thick lines mark the fiducial target sample used in this work. \label{fig:bias_maglim} }
\end{figure*}

\subsection{Dependence of target and comparison with human flags}
\label{sec:depth}

Previously, we have assumed that our weak lensing source sample, or the target sample, is a $i<23.4$ magnitude-limited sample. 
We also noted in \Tref{tab:target_sample} that this target sample is slightly deeper than the DES Y1 weak lensing sample. 
Here we explore a more general situation and ask how our results change when we vary the magnitude limit of the sample.
Effectively this shows, with the same 
spectroscopic sample, the change in the redshift bias from spectroscopic incompleteness as a function of the survey depth. 
The left panels of \Fref{fig:bias_maglim} show the redshift biases for the raw and weighted samples at different depths, for the 
highest source bin in \Fref{fig:bias_main}. As expected, the bias becomes greater as we go to deeper target samples. While the 
reweighting can correct for some of the incompleteness, once the target 
sample is deeper than $i\sim23$, the limitation of the reweighting becomes apparent, as the number of spectra at those depths 
become relatively sparse as well. 

We also show, in the right panels of \Fref{fig:bias_maglim}, the same plot as the left panels, but using human redshift flags instead of 
the RF flags. There are several characteristics of this plot which are different from the left panels that come from the construction of 
two sets of flags: First, the human flags are much fewer in number and therefore noisier. Second, the human flags are more quantised, 
which is expected since the human flags cluster more around integers and the RF flags smooth out this behaviour. Third, the roll-off 
of the curves going from flag 3 to flag 4 in the RF flags seem a lot faster than the human flags. This could be the RF interpolating 
over regions without a lot of data. Fourth, there is one curve ($i\sim23.2$) in the human flags that appears qualitatively different from 
the others; this is likely due to noise. 

We note that, despite these differences, our main conclusion holds for both human and RF flags: at the relevant flag thresholds of 3-3.5, 
the redshift bias in a DES Y1-like weak lensing sample at redshift $>0.9$ exceeds the uncertainty in other calibration methods.  
These results are consistent with that shown in \citet{Bonnett2016} and \citet{Gruen2017}. The former estimated a bias of 
0.05 for the DES Science Verification dataset, while the latter measured the bias in the mean redshift introduced 
from spectroscopic incompleteness in existing data to be at the level of $\sim 0.1$ with a much deeper sample $i<25$. In Appendix~\ref{sec:Gruen} we compare our results with those estimated by \citet{Gruen2017} for a sample matched in limiting i-band magnitude.

\subsection{Discussion}
\label{sec:probabilistic}

The effect that we have investigated can be summarised as a deficit in our knowledge of $p(z, T, {\bf flux})$. Here, {\bf flux} represents the vector of photometry measurements to be used, and clearly correlates with redshift, z, and galaxy SED type, T. Through spectroscopy we have access to only a limited region of this joint probability space of redshift, galaxy SED and flux, while in our target sample we know just the marginal distribution, $p({\bf flux})$. Even in the best cases, e.g. upon completion of the C3R2 programme \citep{Masters2017}, we will still have a selection function in {\bf flux} with respect to the target sample - i.e. there is a co-variate shift. In the samples used in this work this selection function is in both brightness and colour, while in a completed C3R2 it would be in brightness alone. Because $p({\bf flux})$ correlates with z and T in a way that is currently unknowable accurately, matching the marginal distributions in $p({\bf flux})$ of the spectroscopic and target samples cannot guarantee the correct recovery of $p(z, T, {\bf flux})$, or the marginal p(z). In other words, multiple different distributions in $p(z, T, {\bf flux})$ can result in the same marginal distribution, $p({\bf flux})$, and without full knowledge of $p(z, T)$ we are blind to which of them is correct. In practice the situation is likely to be worse than this, and even the eventual C3R2 will likely suffer some degree of (unknown) selection in z and T in addition to {\bf flux}. In order to use a simple weighted spectroscopic sample as an estimator of the redshift distribution of a weak lensing sample we therefore require a sample complete in both colour and amplitude (i.e. {\bf flux}), with very low incompleteness of targeted objects. 

In the case that the spectroscopic samples are not complete (or at least very close to complete) and that failures are not random in redshift, then the fundamental problem that we demonstrate in this paper is formally undefined and therefore cannot be solved. There is no solution that is identifiable in a statistical sense, in that we cannot use the data available to correct for the missing objects. To be clear, we do not know {\em a priori} that a given incomplete spectroscopic sample is inadequate and will introduce a bias, but without filling in the missing data neither can we be confident that we are free of important biases. The approach taken in H18 was to side-step this difficulty through using a calibration sample that was by construction $100\%$ complete, and therefore not subject the sort of selection biases that arise from non-random incompleteness. In doing so the authors made a trade-off, exchanging a possible source of bias for less precise and less accurate redshifts in the form of high-quality photometric redshifts. However, the systematic errors arising from using such high-quality photo-z are more feasibly determined from data than are the effects we have been concerned with in this work. 

While we have explored a very simple algorithm for estimating weak lensing redshift distributions from a spectroscopic sample, the issue of non-identifiability extends to almost all approaches of inferring redshift distributions (one notable exception perhaps being the use of cross-correlation with a reference redshift sample, e.g. \citealt{Newman2008}). For instance, in deriving redshift probabilities via model fitting, we cannot be confident that our SED set is complete if we do not have a complete spectroscopic sample to test them against. Similarly, more sophisticated algorithms that attempt to separate the colour-redshift likelihood from the population density, such as Self Organising Maps (SOMs), may have greater robustness to incompleteness but cannot solve the underlying issue entirely. All we can hope to achieve is to reduce the uncertainties introduced to an acceptable level for a given cosmological analysis. In the next section we introduce a number of possible ways to approach that task.

\section{Mitigation approaches}
\label{sec:mitigation}

We established in the previous section that spectroscopic incompleteness could introduce a bias in the mean redshfit for 
tomographic weak lensing samples. Here we discuss three potential approaches to mitigate such biases. First, we 
consider using lower confidence flags for selection of spectroscopic samples (\Sref{sec:flag2}). Second, we consider 
removing particular regions in colour-magnitude space where the spectra are affected seriously by incompleteness (\Sref{sec:remove_cells}). 
Third, we consider correcting such biases via simulations (\Sref{sec:correct_sims}). 

\subsection{Using lower-confidence redshifts}
\label{sec:flag2}

Incompleteness in spectroscopic samples is not only a challenge for cosmology experiments. The key surveys we 
have been simulating were originally designed to answer questions about the evolution of galaxies and incompleteness 
can bias those answers just as it can bias cosmological parameter estimation. In \cite{Lilly2009} it was suggested 
that the confidence in a spectroscopic redshift could be increased if it is later found to agree with a precise 
photometric redshift (such as are available in the COSMOS field), because the photometric redshift uses 
complementary information. In this way, they proposed a statistically complete sample which supplemented the 
high-confidence redshifts with objects that showed just such agreement.  

From \Fref{fig:bias_main}, it is clear that using objects with confidence flags as low as 2 would achieve a level of bias 
close to the precision quoted in H18, but would be insufficient for the highest tomographic bin in any future analysis. 
We would need to use essentially all galaxies in the sample for spectroscopy to be a useful addition to the validation 
process. It is not obvious that all low-confidence objects will match their respective high-quality photo-$z$, and the worry that some of these are nevertheless wrong is a concern. \cite{Cunha2014} show that even just $2\%$ of galaxies having wrong redshifts will bias the Dark Energy equation of state parameter, $w$, by more than $10\%$. Our simulations are not realistic enough to assess this point in detail, as the fraction of wrong redshift assignments is lower than estimated in real spectroscopic data sets. We leave it to future work, using more sophisticated simulations, to explore this avenue.

\subsection{Removing troublesome regions of colour-magnitude space}
\label{sec:remove_cells}

When probing the large-scale structure, one of the main differences between using weak lensing and using galaxy densities is that 
for weak lensing the galaxy sample used for shear measurements does not need to be complete, since they are 
merely probes of the lensing field. That is, we could attempt to trade statistical precision (i.e. use fewer galaxies) to 
reduce biases caused by systematic errors in the galaxy sample, such as the effect studied in this paper -- 
incompleteness of the spectroscopic samples. A natural approach is to use galaxy photometry to isolate subsets of 
our sample that are likely to be impacted by biases due to incompleteness and exclude them from the 
weak lensing sample. We investigate here one approach for doing this -- using Self-Organising Maps \citep[SOM,][]{Kohonen1982}. 
An SOM maps a high-dimensional space into lower dimensionality (typically 2-D) via an artificial 
neural network; SOMs were introduced as a tool for exploring colour-space coverage of spectroscopic data sets in 
cosmology by \cite{Masters2015}. An SOM enables one to cleanly assign each galaxy to a subsample in the quantised 
photometric space.

We construct our SOM using $g-r$, $r-i$, $i-z$ and $i$-band magnitude from the $i<23.4$ simulated photometric data. We choose a 28$\times$28 square map, with sigma=3, learning rate=0.4 and periodic boundaries, using the python package \texttt{MiniSom}\footnote{https://github.com/JustGlowing/minisom}. We have confirmed that our conclusions are insensitive to reducing the map size and to the precise values of the hyperparameters. However for this simple exploration we have not attempted to optimise the SOM parameters. 
For each cell in the SOM, we collect the spectroscopic redshifts of the galaxies that belong to that cell and determine 
whether to retain or discard that cell based on a simple metric. We perform two runs, once with each of the following criteria:
\begin{itemize}
\item If the spectroscopic failure rate is above $f_{\rm fail}$ (we use Flag $<3$ as a criteria for a failed redshift)
\item If the intrinsic true redshift dispersion in the cell is greater than $\sigma_{\rm spec}$ 
\end{itemize} 

After discarding certain SOM cells in both the spectroscopic sample and the weak lensing sample, we reweight the 
spectroscopic galaxies to match the weak lensing sample in the same way as described in \Sref{sec:reweighting}. 
We then explore how the bias due to spectroscopic incompleteness changes as we removed progressively more cells 
from the analysis.

\begin{figure}
\begin{center}
\includegraphics[width=\columnwidth]{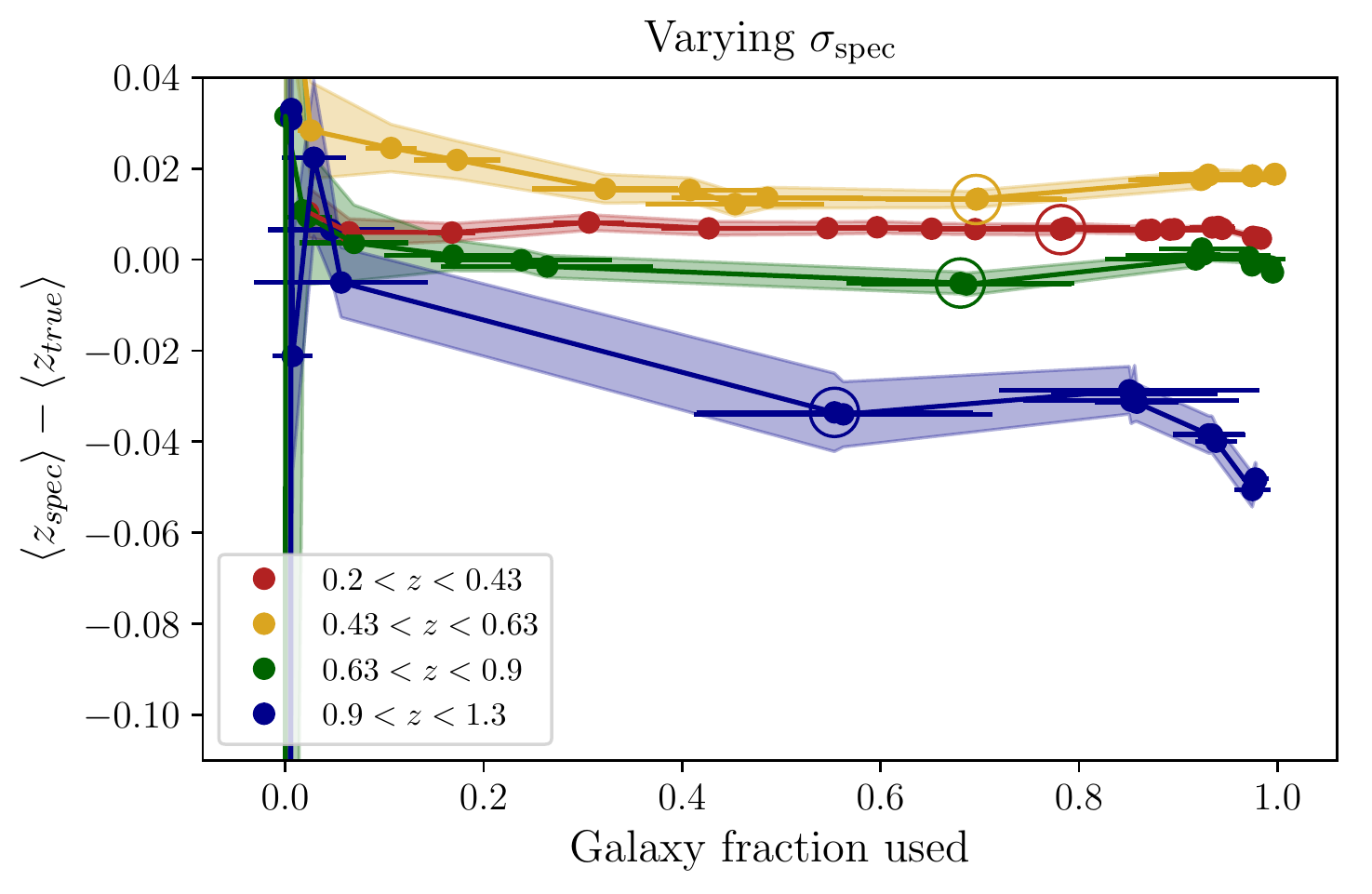}
\includegraphics[width=\columnwidth]{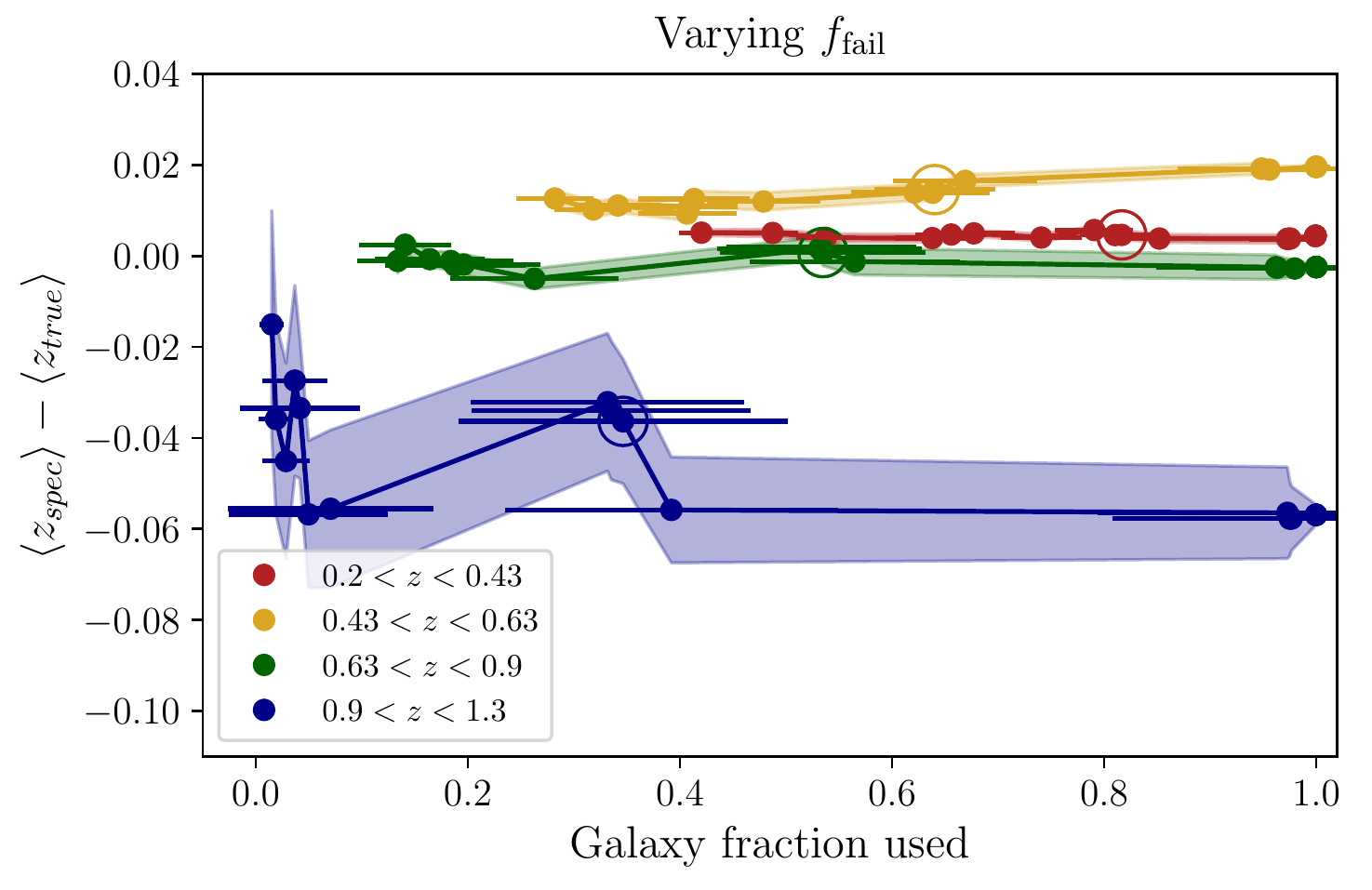}
\end{center}
\caption[]{Redshift bias from spectroscopic incompleteness in the four weak lensing tomographic bins as a function of 
the fraction of source galaxies used. All values here are after reweighting has been applied. From right to left, we remove increasingly 
more cells from an SOM quantisation of the colour-magnitude space. The removal of cells for the upper (lower) panel is based on the intrinsic spectroscopic redshift 
scatter $\sigma_{\rm spec}$ (fraction of galaxies where we failed to get a high-confidence redshfit $f_{\rm fail}$). Large open circles mark $\sigma_{\rm spec}=0.2$ (upper panel) and $f_{\rm fail}=0.5$ (lower panel), the mid points of the ranges we consider.  \label{fig:som}}
\end{figure}

\Fref{fig:som} shows the bias in each redshift bin as a function of the fraction of the $i<23.4$ photometric sample that is 
retained for the case where the two different criteria were used to discard SOM cells. We have also explored the extreme case of the effect, using the $i<24$ sample, to determine whether it is a suitable approach that could be used for the final DES data set. Our conclusions are unaltered with respect to the $i<23.4$ sample, but reflect the larger bias seen in \Fref{fig:bias_maglim}. The uncertainties were derived by 50 realisations of the SOM random seed, and therefore represent the uncertainty in the SOM method only.

It is clear from \Fref{fig:som} that redshift dispersion is the more effective indicator to use for overcoming biases from spectroscopic incompleteness. However, to reduce the impact to a similar level of uncertainty as 
H18 we must remove on the order of two thirds of our highest redshift bin. 
Therefore, while it is a possible mitigation strategy, this seems impractical for use in DES -- the degradation of statistical power will have a much larger impact than simply marginalising over a few percent of redshift biases from the spectroscopic incompleteness. Note, however, that as we reduce statistical precision, the impact of photo-z biases in the final parameter constraints will become less important. With poorer statistical power we would be able to allow a greater error budget in the redshift distributions and could therefore optimise the sample with somewhat greater numbers than \Fref{fig:som} suggests.

This is not to say that approaches aiming to reduce incompleteness-related biases with SOMs are necessarily futile. At higher dimensionality (e.g. KiDS+VIKING, Euclid+EXT, DES+VIDEO) they may see greater success, as 
the SOM cells will have narrower intrinsic redshift dispersion (provided photometric uncertainties are not large). Moreover, implementations with greater sophistication than the fairly naive one used in this work \citep[e.g.][]{Alarcon2019,Buchs2019,Sanchez2019,Wright2019} could be capable of more promising results. Most of these methods seek to gain an advantage by also utilising higher-dimensional data. However, these are clearly areas of development work that are beyond the scope of this paper.

\subsection{Using simulation results to correct biases in real samples}
\label{sec:correct_sims}

It is tempting to ask whether we could simply use the bias computed from our simulations (or a more 
sophisticated version of them) to correct the mean redshift of real spectroscopic data sets, and thereby 
be able to use spectroscopy to validate the redshift distributions in weak lensing experiments. 
It could be worth further investigation, but
it would be extremely challenging in the the short term. 
To accurately simulate the bias directly we would need to know the true galaxy distribution in redshift and 
SED type, have accurate galaxy SEDs and be able to select a target weak lensing sample from the 
simulations that represents the true target sample. Clearly, if we were confident that we had all of this 
information then our problem is already solved and we would have no need to perform these simulations.

One could instead imagine deriving galaxy type and redshift-dependent incompleteness factors, which 
removes the need for an a priori correct redshift distribution or mix of galaxy types. As the redshift 
confidence flag depends on feature strength in the spectra, we would still need to be confident that the 
equivalent widths of emission and absorption lines in our simulations are appropriate at any given redshift. 
Though our knowledge of high-redshift galaxy spectra and SEDs has increased greatly over recent years \citep[e.g.][]{Maltby2016,Wuyts2016,Kashino2017,Forrest2018}, 
we are still not quite at that stage.

\section{Summary}
\label{sec:conclusion}

One of the crucial components for weak lensing as a cosmological probe is the redshift 
distribution of the source galaxy sample. Even small biases in the estimated redshift distributions can lead to important biases in the final cosmological parameter estimates. These potential biases in redshift, and the methods to overcome them or incorporate them in analyses via nuisance parameters, have been the focus of a growing literature \citep[e.g.][]{Cunha2012,Masters2015,Bonnett2016,Gruen2017,Hoyle2018,Joudaki2019}. 
In this work, we examine one particular method in which the redshift distributions are obtained -- by directly 
weighting the redshifts of a spectroscopic sample based on the photometric 
properties of the spectroscopic sample and the weak lensing source galaxies. In an idealised situation where 
the following is true, this method results in an unbiased estimate of the redshift distribution up to the limit of sample variance:
\begin{itemize}
\item The spectroscopic redshifts of the sample being weighted are all correct.
\item The uncertainties in the photometry of the spectroscopic sample are representative of the target sample.
\item At any given locale in photometric space, the available spectroscopic redshifts are equivalent to a random draw from the true redshift distribution of the target sample in that same locale.
\end{itemize} 

We examine the validity of the last assumption here through the use of simulated spectroscopic surveys, ensuring the other two conditions are met by construction. In particular, we investigate how the spectroscopic samples that are assembled for this purpose are typically incomplete, either due to the imposed survey selection function or due to the ``redshifting" procedure -- human observers inspecting each spectrum and assigning confidence flags to indicate how secure the determined redshift is. We show that targetting strategies that include a band or bands unavailable to the weak lensing source galaxies, such as the one employed for the VIPERS data set, can introduce biases of up to $\Delta~z\sim0.04$ ($\Delta~z\sim0.015$ for the case most closely reflecting the spectroscopy available to DES). While this result is specific to the case of combining VIPERS and VVDS Wide in varying amounts, any spectroscopic survey using a target selection function outside of the source galaxy photometric space could result in a non-negligible bias.

As only highly-confident redshifts can be used to construct or validate the desired source galaxy redshift distributions, the redshifting process introduces a subtle selection effect that is analogous to that caused by the aforementioned targetting strategies. Spectra of certain types and redshifts are preferentially given a lower flag values due to having fewer or less prominent features such as spectral lines or breaks. These features are not uniquely determinable from the broad band photometry available to weak lensing experiments, and could lead to a redshift-dependent success rate in determining a confident redshift at fixed locale in colour-magnitude space. In this way, an incompleteness in the spectroscopic sample could lead to a different inferred mean 
redshift of the target sample compared to the case of having a complete spectroscopic sample. This incompleteness-related bias from the spectroscopic sample cannot be removed with the commonly-employed re-weighting procedure of \citet{Lima2008}.

We carry out a simulated analysis to estimate the order of magnitude of this source of bias for a dataset 
similar to the first year weak lensing analysis from the Dark Energy Survey (DES). Simulated 
spectra (which include simple noise and sky models) are constructed to match three of the four key spectroscopic surveys covered by DES and passed to experienced 
observers, or ``redshifters", to assign quality flags. Using these human-redshifted spectra as a training 
set, we use a random forest approach to enlarge the sample of spectra to a similar sample size as 
that available to be used in DES Y1. Next, we derive the re-weighted redshift distributions for the 
DES Y1-like weak lensing sample and compare the mean redshift of these distributions to that of the true 
redshift distribution. We find that at a conservative redshift flag threshold, Flag $\ge3$, the incompleteness-induced biases, $z_{\rm spec}-z_{\rm true}$, in mean redshift are 0.004, 0.020, -0.002 and -0.057 for our four tomographic bins ($0.2<z<0.43$, $0.43<z<0.63$, $0.63<z<0.9$ and $0.9<z<1.3$, respectively). Using only human-redshifted objects, these become 0.005, 0.018, -0.003 and -0.045 respectively.

We further explore how the bias in the highest tomographic bin depends on the magnitude limit of the target sample, finding that it becomes rapidly and progressively worse at depths greater than DES-Y1. In two of our bins, these biases are at a similar or larger level than the 
uncertainties of the photometric redshift derived from COSMOS15, suggesting that 
direct re-weighting of the spectroscopic redshifts is not an appropriate approach for DES Y1, nor for the future DES analyses that will require still greater accuracy.

The impact of incompleteness is likely to be most severe for surveys in which the intrinsic redshift distribution is broad in at least part of the photometric space (either due to low dimensionality or significant photometric errors), and where the target sample extends to redshifts that are challenging to recover with current high-multiplex spectrographs -- the so-called ``redshift desert''. Stage IV weak lensing experiments such as Euclid and LSST are therefore likely to face great difficulties in using spectroscopic redshifts for direct validation, not withstanding tremendous efforts such as the C3R2 programme \citep{Masters2017, Masters2019}. 

It is worth noting that spectroscopic incompleteness is not only problematic for direct calibration of photometric redshift distributions. Template SEDs used in deriving individual galaxy PDFs are either drawn from low-redshift data, where obtaining a representative sample to low luminosity is possible, or synthetic composite SEDs build up from stellar isochrones. To be useful at the accuracy required from cosmology analyses, these SEDs and their associated prior probabilities need to be calibrated across the redshift range that will be used or, perhaps, jointly estimated along with photometric redshifts \citep{Leistedt2019}. Spectroscopic redshifts are frequently used for this purpose (though low dispersion spectra or precise photometric redshift might also be used, \citealt{Forrest2018, Hoyle2018}). Redshift-dependent selection effects may therefore subtly distort the calibrations applied, with the risk that these distortions too introduce redshift biases. 

There are, however, potential ways one could remove the redshift biases introduced in a DIR-like method by incomplete spectroscopic samples. We showed in \Sref{sec:remove_cells} that the bias due to incompleteness could be substantially reduced by excluding regions of photometric space from a weak lensing analysis, but at the cost of removing $60-70\%$ of the target sample in the highest redshift interval. The situation could greatly improve if a larger number of bands were available. Amongst the current methods being developed to deliver robust redshift distributions, perhaps the most encouraging are the combination of photometric and clustering information \citep{Alarcon2019,Rau2019} or methods of inferring the intrinsic galaxy population by forward modelling the entire survey transfer function onto simulated observed skies \citep{Herbel2017, Fagioli2018}. 

Finally, we note that there are a number of simplifications that were used in this analysis (as summarised in 
\Sref{sec:simplifications}) as accounting for all the details of the different spectroscopic samples 
in the simulations is prohibitively impractical. Our approach likely produces more idealised 
spectra and thus a conservative estimate of the bias in redshift. We expect the true incompleteness  
in spectroscopic samples is equal to or worse than that which we have found, which implies a systematic uncertainty 
larger than what can be tolerated in  
present and future lensing surveys.  

\section*{Acknowledgements}

We thank Andrina Nicola for early consultant on spectra simulations. CC is supported by the Henry Luce Foundation. 

Funding for the DES Projects has been provided by the U.S. Department of Energy, the U.S. National Science Foundation, the Ministry of Science and Education of Spain, 
the Science and Technology Facilities Council of the United Kingdom, the Higher Education Funding Council for England, the National Center for Supercomputing 
Applications at the University of Illinois at Urbana-Champaign, the Kavli Institute of Cosmological Physics at the University of Chicago, 
the Center for Cosmology and Astro-Particle Physics at the Ohio State University,
the Mitchell Institute for Fundamental Physics and Astronomy at Texas A\&M University, Financiadora de Estudos e Projetos, 
Funda{\c c}{\~a}o Carlos Chagas Filho de Amparo {\`a} Pesquisa do Estado do Rio de Janeiro, Conselho Nacional de Desenvolvimento Cient{\'i}fico e Tecnol{\'o}gico and 
the Minist{\'e}rio da Ci{\^e}ncia, Tecnologia e Inova{\c c}{\~a}o, the Deutsche Forschungsgemeinschaft and the Collaborating Institutions in the Dark Energy Survey. 

The Collaborating Institutions are Argonne National Laboratory, the University of California at Santa Cruz, the University of Cambridge, Centro de Investigaciones Energ{\'e}ticas, 
Medioambientales y Tecnol{\'o}gicas-Madrid, the University of Chicago, University College London, the DES-Brazil Consortium, the University of Edinburgh, 
the Eidgen{\"o}ssische Technische Hochschule (ETH) Z{\"u}rich, 
Fermi National Accelerator Laboratory, the University of Illinois at Urbana-Champaign, the Institut de Ci{\`e}ncies de l'Espai (IEEC/CSIC), 
the Institut de F{\'i}sica d'Altes Energies, Lawrence Berkeley National Laboratory, the Ludwig-Maximilians Universit{\"a}t M{\"u}nchen and the associated Excellence Cluster Universe, 
the University of Michigan, the National Optical Astronomy Observatory, the University of Nottingham, The Ohio State University, the University of Pennsylvania, the University of Portsmouth, 
SLAC National Accelerator Laboratory, Stanford University, the University of Sussex, Texas A\&M University, and the OzDES Membership Consortium.

Based in part on observations at Cerro Tololo Inter-American Observatory, National Optical Astronomy Observatory, which is operated by the Association of 
Universities for Research in Astronomy (AURA) under a cooperative agreement with the National Science Foundation.

The DES data management system is supported by the National Science Foundation under Grant Numbers AST-1138766 and AST-1536171.
The DES participants from Spanish institutions are partially supported by MINECO under grants AYA2015-71825, ESP2015-66861, FPA2015-68048, SEV-2016-0588, SEV-2016-0597, and MDM-2015-0509, 
some of which include ERDF funds from the European Union. IFAE is partially funded by the CERCA program of the Generalitat de Catalunya.
Research leading to these results has received funding from the European Research
Council under the European Union's Seventh Framework Program (FP7/2007-2013) including ERC grant agreements 240672, 291329, and 306478.
We  acknowledge support from the Brazilian Instituto Nacional de Ci\^encia
e Tecnologia (INCT) e-Universe (CNPq grant 465376/2014-2).

This manuscript has been authored by Fermi Research Alliance, LLC under Contract No. DE-AC02-07CH11359 with the U.S. Department of Energy, Office of Science, Office of High Energy Physics.

\section*{Data Availability}

The data underlying this article were accessed from Kavli Institute for Particle Astrophysics and Cosmology. The derived data generated in this research will be shared on reasonable request to the corresponding author.

\setlength{\bibhang}{2.0em}
\setlength\labelwidth{0.0em}
\bibliography{SpecIncompl}
\bibliographystyle{mn2e_2author_arxiv.bst}

\section*{Affiliations}
$^{1}$ Department of Physics \& Astronomy, University College London, Gower Street, London, WC1E 6BT, UK\\
$^{2}$ D\'{e}partement de Physique Th\'{e}orique and Center for Astroparticle Physics, Universit\'{e} de Gen\`{e}ve, 24 quai Ernest Ansermet, CH-1211 Geneva, Switzerland\\
$^{3}$ Department of Physics, ETH Zurich, Wolfgang-Pauli-Strasse 16, CH-8093 Zurich, Switzerland\\
$^{4}$ Department of Astronomy and Astrophysics, University of Chicago, Chicago, IL 60637, USA\\
$^{5}$ Kavli Institute for Cosmological Physics, University of Chicago, Chicago, IL 60637, USA\\
$^{6}$ Centro de Investigaciones Energ\'eticas, Medioambientales y Tecnol\'ogicas (CIEMAT), Madrid, Spain\\
$^{7}$ School of Mathematics and Physics, University of Queensland,  Brisbane, QLD 4072, Australia\\
$^{8}$ Max Planck Institute for Extraterrestrial Physics, Giessenbachstrasse, 85748 Garching, Germany\\
$^{9}$ Universit\"ats-Sternwarte, Fakult\"at f\"ur Physik, Ludwig-Maximilians Universit\"at M\"unchen, Scheinerstr. 1, 81679 M\"unchen, Germany\\
$^{10}$ Department of Physics, Stanford University, 382 Via Pueblo Mall, Stanford, CA 94305, USA\\
$^{11}$ Kavli Institute for Particle Astrophysics \& Cosmology, P. O. Box 2450, Stanford University, Stanford, CA 94305, USA\\
$^{12}$ SLAC National Accelerator Laboratory, Menlo Park, CA 94025, USA\\
$^{13}$ Laborat\'orio Interinstitucional de e-Astronomia - LIneA, Rua Gal. Jos\'e Cristino 77, Rio de Janeiro, RJ - 20921-400, Brazil\\
$^{14}$ Observat\'orio Nacional, Rua Gal. Jos\'e Cristino 77, Rio de Janeiro, RJ - 20921-400, Brazil\\
$^{15}$ The Research School of Astronomy and Astrophysics, Australian National University, ACT 2601, Australia\\
$^{16}$ Australian Astronomical Optics, Macquarie University, North Ryde, NSW 2113, Australia\\
$^{17}$ Lowell Observatory, 1400 Mars Hill Rd, Flagstaff, AZ 86001, USA\\
$^{18}$ McWilliams Center for Cosmology, Department of Physics, Carnegie Mellon University, Pittsburgh, PA 15213, USA\\
$^{19}$ Department of Astronomy, University of California, Berkeley,  501 Campbell Hall, Berkeley, CA 94720, USA\\
$^{20}$ Santa Cruz Institute for Particle Physics, Santa Cruz, CA 95064, USA\\
$^{21}$ Cerro Tololo Inter-American Observatory, National Optical Astronomy Observatory, Casilla 603, La Serena, Chile\\
$^{22}$ Departamento de F\'isica Matem\'atica, Instituto de F\'isica, Universidade de S\~ao Paulo, CP 66318, S\~ao Paulo, SP, 05314-970, Brazil\\
$^{23}$ Fermi National Accelerator Laboratory, P. O. Box 500, Batavia, IL 60510, USA\\
$^{24}$ Instituto de Fisica Teorica UAM/CSIC, Universidad Autonoma de Madrid, 28049 Madrid, Spain\\
$^{25}$ Department of Physics and Astronomy, University of Pennsylvania, Philadelphia, PA 19104, USA\\
$^{26}$ CNRS, UMR 7095, Institut d'Astrophysique de Paris, F-75014, Paris, France\\
$^{27}$ Sorbonne Universit\'es, UPMC Univ Paris 06, UMR 7095, Institut d'Astrophysique de Paris, F-75014, Paris, France\\
$^{28}$ Jodrell Bank Center for Astrophysics, School of Physics and Astronomy, University of Manchester, Oxford Road, Manchester, M13 9PL, UK\\
$^{29}$ Department of Astronomy, University of Illinois at Urbana-Champaign, 1002 W. Green Street, Urbana, IL 61801, USA\\
$^{30}$ National Center for Supercomputing Applications, 1205 West Clark St., Urbana, IL 61801, USA\\
$^{31}$ Institut de F\'{\i}sica d'Altes Energies (IFAE), The Barcelona Institute of Science and Technology, Campus UAB, 08193 Bellaterra (Barcelona) Spain\\
$^{32}$ Institut d'Estudis Espacials de Catalunya (IEEC), 08034 Barcelona, Spain\\
$^{33}$ Institute of Space Sciences (ICE, CSIC),  Campus UAB, Carrer de Can Magrans, s/n,  08193 Barcelona, Spain\\
$^{34}$ Physics Department, 2320 Chamberlin Hall, University of Wisconsin-Madison, 1150 University Avenue Madison, WI  53706-1390\\
$^{35}$ INAF-Osservatorio Astronomico di Trieste, via G. B. Tiepolo 11, I-34143 Trieste, Italy\\
$^{36}$ Institute for Fundamental Physics of the Universe, Via Beirut 2, 34014 Trieste, Italy\\
$^{37}$ Department of Physics, IIT Hyderabad, Kandi, Telangana 502285, India\\
$^{38}$ Faculty of Physics, Ludwig-Maximilians-Universit\"at, Scheinerstr. 1, 81679 Munich, Germany\\
$^{39}$ Department of Astronomy, University of Michigan, Ann Arbor, MI 48109, USA\\
$^{40}$ Department of Physics, University of Michigan, Ann Arbor, MI 48109, USA\\
$^{41}$ Center for Cosmology and Astro-Particle Physics, The Ohio State University, Columbus, OH 43210, USA\\
$^{42}$ Department of Physics, The Ohio State University, Columbus, OH 43210, USA\\
$^{43}$ Center for Astrophysics $\vert$ Harvard \& Smithsonian, 60 Garden Street, Cambridge, MA 02138, USA\\
$^{44}$ Department of Astronomy/Steward Observatory, University of Arizona, 933 North Cherry Avenue, Tucson, AZ 85721-0065, USA\\
$^{45}$ George P. and Cynthia Woods Mitchell Institute for Fundamental Physics and Astronomy, and Department of Physics and Astronomy, Texas A\&M University, College Station, TX 77843,  USA\\
$^{46}$ Department of Astrophysical Sciences, Princeton University, Peyton Hall, Princeton, NJ 08544, USA\\
$^{47}$ Instituci\'o Catalana de Recerca i Estudis Avan\c{c}ats, E-08010 Barcelona, Spain\\
$^{48}$ School of Physics and Astronomy, University of Southampton,  Southampton, SO17 1BJ, UK\\
$^{49}$ Brandeis University, Physics Department, 415 South Street, Waltham MA 02453\\
$^{50}$ Computer Science and Mathematics Division, Oak Ridge National Laboratory, Oak Ridge, TN 37831\\
$^{51}$ Department of Physics, Duke University Durham, NC 27708, USA\\
$^{52}$ Department of Physics and Astronomy, Pevensey Building, University of Sussex, Brighton, BN1 9QH, UK\\

\appendix{}

\section{Features used for the RF training}
\label{sec:features}

In \Tref{tab:features} we list the spectroscopic features used in the RF training described in \Sref{sec:redshift}.
\begin{table}
\begin{center}
\caption{Features used for the RF training.}
\begin{tabular}{cc}
\hline
Wavelength ($\rm \AA$) & Feature \\ \hline
1215.7 & Ly$\alpha$ \\
240.0 & NV \\
1303.0 & OI \\
1334.5 & CII \\
1397.0 & SiIV1393+OIV1402 \\
1549.0 & CIV1548 \\
1640.0 & HeII \\
1909.0 & CIII] \\
2142.0 & NII] \\
2626.0 & FeII \\
2799.0 & MgII \\
2852.0 & MgI \\
2964.0 & FeII \\ 
3727.5 & [OII] \\
3933.7 & CaK \\
3968.5 & CaH \\
4101.7 & H$\delta$ \\
4304.4 & Gband \\
4340.4 & H$\gamma$ \\
4861.3 & H$\beta$ \\
4958.9 & [OIIIa] \\
5006.8 & [OIIIb] \\
5175.0 & MgI \\
5269.0 & CaFe \\
5711.0 & MgI \\
6562.8 & H$\alpha$ \\
6725.0 & [SII]6717.0+6731.3\\
\hline
\end{tabular}
\label{tab:features}
\end{center}
\end{table}

\section{Effect of field-to-field variance: test on VVDS Deep}
\label{sec:vvdsdeep}

\begin{figure}
\begin{center}
\includegraphics[width=\columnwidth]{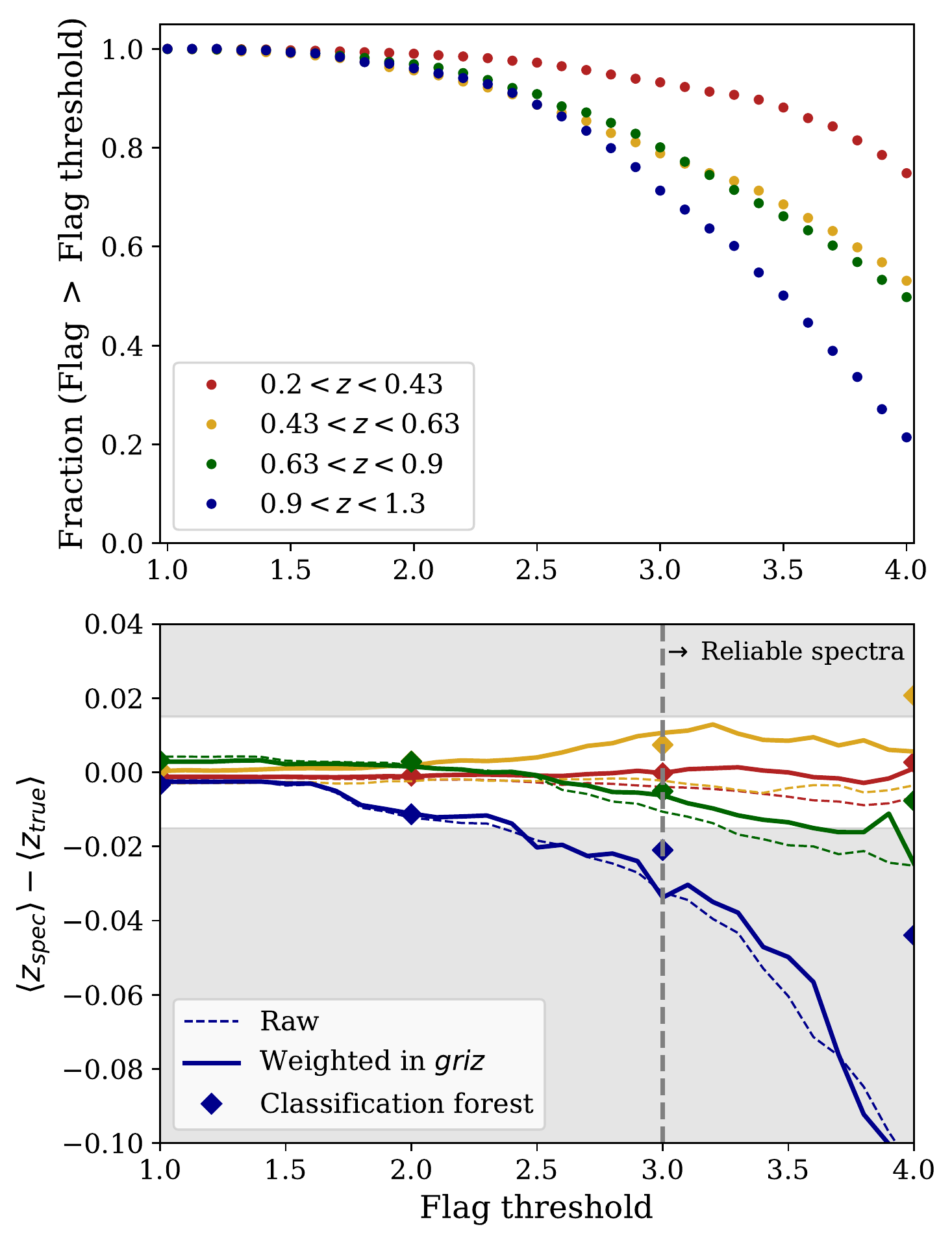}
\end{center}
\caption[]{Same as the top two panels of \Fref{fig:bias_main}, but for VVDS Deep (field 1) alone. \label{fig:bias_vvdsdeep}
}
\end{figure}

In order to remove the impact of field-to-field variance seen in \Sref{sec:results} and isolate the bias in the mean 
redshift purely due to spectroscopic incompleteness, we repeat the analysis using only the main VVDS Deep field 
(containing 12,932 objects) and a target sample co-located on the sky. The results are shown in \Fref{fig:bias_vvdsdeep}, 
where the lines and symbols have the same meanings as the upper two panels in \Fref{fig:bias_main}.   

We note that the general shape of the curves in \Fref{fig:bias_vvdsdeep} is qualitatively different from \Fref{fig:bias_main}. 
For the VVDS Deep sample, the bias in mean redshift before re-weighting is much smaller at low flag thresholds, reflecting the 
reduced field-to-field variance. The final bias in redshifts after reweighting, however, does not differ significantly from the 
previous results where the full sample is used. This illustrates 
that in general, field-to-field variance could largely be corrected for through the re-weighting process (see also H18).   

\section{Comparison with Gruen $\&$ Brimioulle (2017)}
\label{sec:Gruen}

In order to compare our results on selection bias to real spectroscopic samples, we run an analysis of the type of \citet{Gruen2017}. We use the same catalogues and formalism but with a magnitude cut to match the DES-like sample used in this paper, and with four redshift bins. We do not force the redshift bins to be the same as those in the DES Y1 analysis, instead allowing the bin boundaries in redshift to adapt so that there are equal numbers of spectroscopic objects in each, as was done in \citet{Gruen2017}.

The catalogue used in \citet{Gruen2017} was constructed from the overlap of the four CFHTLS Deep fields\footnote{\texttt{http://www.cfht.hawaii.edu/Science/CFHLS/ cfhtlsdeepwidefields.html}} with near-infrared imaging from the WIRCam Deep Survey (WIRDS, \citealt{Bielby2012}), which we cut at $i_{\rm CFHT}<23.4$. Photometric redshifts are determined by a template fit with Photo-Z \citep{Bender2001} to the $ugriz$ CFHT and $JHK_s$ WIRDS fluxes. This is the only source of redshift we use in this test, which assumes that the 8-band photometric redshift closely approximates the truth. Typical uncertainties on photo-$z$ from 8-bands covering the $u$ to $K$-band wavelength range are $\sim 3-4\%$ at the depths considered here. See, for instance, \citet{Hartley2013}.

Spectroscopic redshift measurements are compiled from the VIMOS VLT Deep Survey (VVDS-Deep, \citealt{LeFevre2005}), the VIMOS Public Extragalactic Survey (VIPERS, \citealt{Garilli2014,Guzzo2014}), the VIMOS Ultra Deep Survey (VUDS, \citealt{LeFvre2015,Tasca2017}), zCOSMOS-bright and zCOSMOS-deep \citep{Lilly2007}, or the Deep Extragalactic Evolutionary Probe-2 (DEEP2) survey \citep{Newman2013}. We mark objects with Flag 3 or 4 as successful spectroscopic redshift determinations and label these as ``spectroscopically selected''. Note that VIPERS, zCOSMOS-deep, VVDS and DEEP2 have color-based pre-selection of targets applied (cf. \Sref{sec:vipers}) in addition to the purely spectroscopic selection effects primarily studied in this work (\Sref{sec:results}).

\begin{figure}
\begin{center}
\includegraphics[width=\columnwidth]{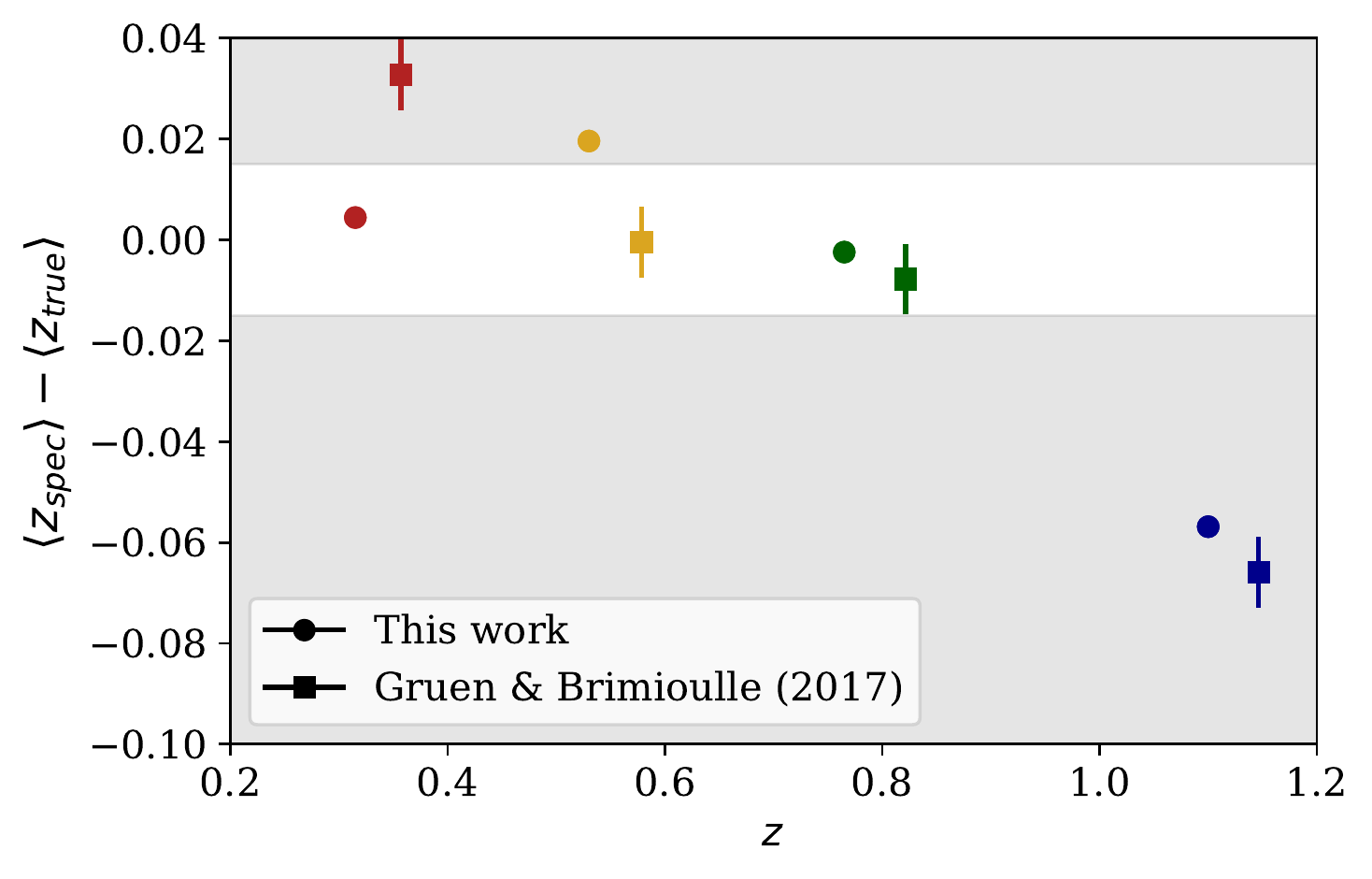}
\end{center}
\caption[]{Comparison of the bias in mean redshift found through our simulated data sets in this work and the observed data sets used in \citet{Gruen2017}. \label{fig:Gruen}
}
\end{figure}

We build a single colour-magnitude decision tree by performing splits at the median of whichever of $i,g-i,r-i$ or $i-z$ separates the two subsamples best in redshift. These subsamples, called leaves, are divided further until no additional split significantly separates the subsamples in redshift, or until there are fewer than 10 spectroscopically selected galaxies left in a leaf. These leaves are then ordered by the mean photometric redshift of all galaxies they contain and separated into bins of consecutive leaves such that each contains approximately one quarter of the total number of photometric galaxies. The mean true redshift of each bin is defined as the mean photometric redshift of all photometric galaxies it contains. We make a second estimate of the mean redshift using the photometric redshifts of the spectroscopically selected objects. When computing this second mean, each spectroscopically-selected galaxy is weighted by the ratio of photometric to spectroscopically-selected galaxies in its leaf. In this way, we emulate reweighting by $griz$ colour-magnitude to reduce the spectroscopic selection bias.

\Fref{fig:Gruen} shows the difference of these two estimates, i.e.~the spectroscopic selection bias, as a function of mean photometric redshift of the bin. Unlike the main result of this paper, this is a mix of the VIMOS-like implicit selection effect of \Sref{sec:results} with corresponding effects for DEIMOS/DEEP2 and color pre-selection effects as in \Sref{sec:vipers}. Depending on the mix of spectroscopic surveys used, the spectroscopic selection bias found could potentially vary substantially. Despite these differences, the overall amplitude and redshift trend of spectroscopic selection bias is rather similar to the main result of this work. 

\section{Reweighted colour space}
\label{sec:weight_plots}

\begin{figure*}
\begin{center}
  \includegraphics[width=\textwidth]{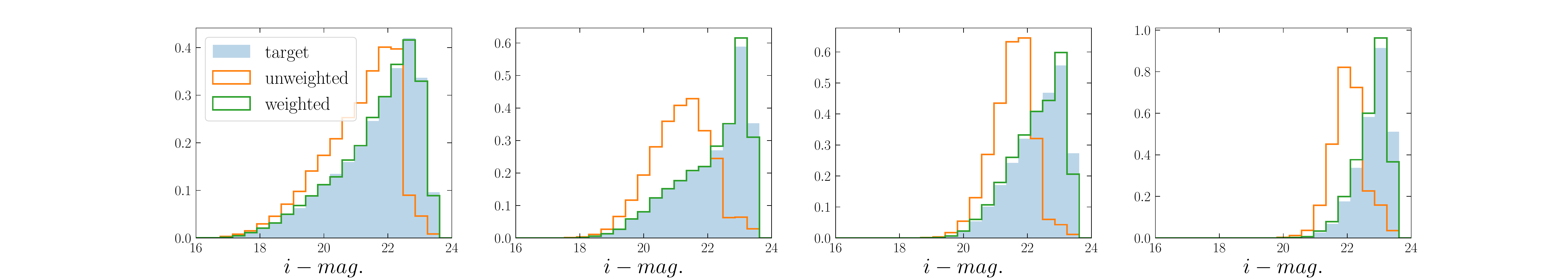}
  \includegraphics[width=\textwidth]{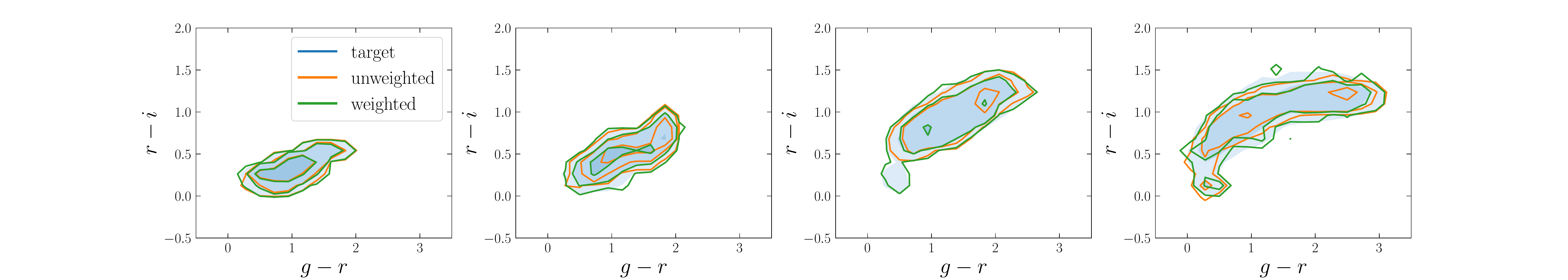}
  \includegraphics[width=\textwidth]{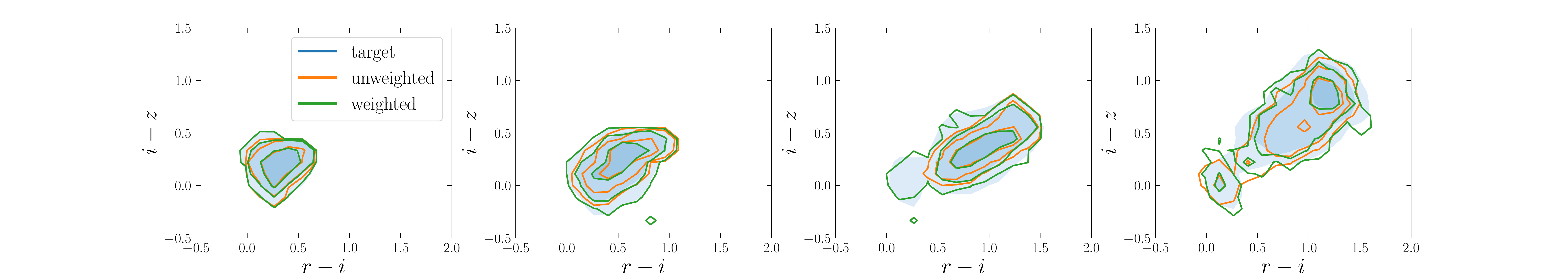}
\end{center}
\caption[]{Weighted (green) and unweighted (orange) magnitude and colour-space distributions in our four redshift intervals, compared with the target photometric sample. Distributions are shown for the fiducial case: $i<23.4$ and spectroscopic Flag~$\ge3$. The weighted spectroscopic sample closely mimics the target photometric sample in terms of their photometric distributions, correcting the mismatch in sampling from the unweighted incomplete spectroscopic sample. Nevertheless, the redshift distribution is not correctly recovered (see \Sref{sec:results}). \label{fig:col_weight}
}
\end{figure*}

One very basic requirement of the reweighting scheme used in \citet{Lima2008} is that the entire colour-magnitude space of the target data set is sampled by objects with spectroscopic redshifts (albeit poorly in some regions). If it is not, then the photometric distributions cannot be matched, and there would be no reason to believe that the resulting redshift distribution would be representative of the target sample. In \Fref{fig:col_weight} we show the photometric space of the target galaxy sample alongside the weighted and unweighted distributions of objects with successful spectroscopic redshift assignments, for our fiducial case: Flag $\ge3$, $i<23.4$, in the simulations used in \Sref{sec:results0}. The weights applied to the spectroscopic data do a good job in replicating the photometric space of the target data set. However, as we showed in \Sref{sec:results}, the redshift distribution is not accurately recovered in all four redshift intervals. 

\end{document}